\documentclass[aps,pra,twocolumn,amsmath,amssymb,nofootinbib,superscriptaddress]{revtex4-2}
\usepackage{todonotes}
\usepackage{times}
\usepackage{graphicx}
\usepackage{dcolumn}
\usepackage{bm}
\usepackage{amsmath}
\usepackage{amsthm}
\usepackage{braket}
\usepackage{indentfirst}
\usepackage[colorlinks]{hyperref}
\usepackage{xcolor}
\usepackage{tabularx}
\usepackage{multirow}
\usepackage{booktabs}
\usepackage{mathtools}

\usepackage[normalem]{ulem}
\usepackage{array}

\usepackage{subfigure}
\usepackage[linesnumbered,ruled,vlined]{algorithm2e}
\usepackage{algpseudocode}
\usepackage{amsmath}
\usepackage{amssymb}
\usepackage{verbatim}
\usepackage{makecell}
\usepackage[normalem]{ulem}

\begin{document}
 
\title{Physics-inspired Machine Learning for Quantum Error Mitigation}

\author{Xiao-Yue Xu}
\thanks{These two authors contributed equally.}
\affiliation{Henan Key Laboratory of Quantum Information and Cryptography, Zhengzhou, Henan 450000, China}
\author{Xin Xue}
\thanks{These two authors contributed equally.}
\affiliation{Department of Computer Science, Beihang University, Beijing 100191}
\affiliation{Beijing Advanced Innovation Center for Big Data and Brain Computing, Beijing, 100191}

\author{Tianyu Chen}
\affiliation{Department of Computer Science, Beihang University, Beijing 100191}
\affiliation{Beijing Advanced Innovation Center for Big Data and Brain Computing, Beijing, 100191}

\author{Chen Ding}
\affiliation{Henan Key Laboratory of Quantum Information and Cryptography, Zhengzhou, Henan 450000, China}

\author{Tian Li}
\affiliation{Henan Key Laboratory of Quantum Information and Cryptography, Zhengzhou, Henan 450000, China}

\author{Haoyi Zhou}
\email{haoyi@buaa.edu.cn}
\affiliation{Department of Software, Beihang University, Beijing 100191}
\affiliation{Beijing Advanced Innovation Center for Big Data and Brain Computing, Beijing, 100191}

\author{He-Liang Huang}
\email{quanhhl@ustc.edu.cn}
\affiliation{Henan Key Laboratory of Quantum Information and Cryptography, Zhengzhou, Henan 450000, China}

\author{Wan-Su Bao}
\email{bws@qiclab.cn}
\affiliation{Henan Key Laboratory of Quantum Information and Cryptography, Zhengzhou, Henan 450000, China}

\date{\today}

\begin{abstract}
\noindent\textbf{Noise is a major obstacle in current quantum computing, and Machine Learning for Quantum Error Mitigation (ML-QEM) promises to address this challenge, enhancing computational accuracy while reducing the sampling overheads of standard QEM methods. Yet, existing models lack physical interpretability and rely heavily on extensive datasets, hindering their scalability in large-scale quantum circuits. To tackle these issues, we introduce the Neural Noise Accumulation Surrogate (NNAS), a physics-inspired neural network for ML-QEM that incorporates the structural characteristics of quantum noise accumulation within multi-layer circuits, endowing the model with physical interpretability. Experimental results demonstrate that NNAS outperforms current methods across a spectrum of metrics, including error mitigation capability, quantum resource consumption, and training dataset size. Notably, for deeper circuits where QEM methods typically struggle, NNAS achieves a remarkable reduction of over half in errors. NNAS also demands substantially fewer training data, reducing dataset reliance by at least an order of magnitude, due to its ability to rapidly capture noise accumulation patterns across circuit layers. 
This work pioneers the integration of quantum process-derived structural characteristics into neural network architectures, broadly enhancing QEM's performance and applicability, and establishes an integrative paradigm that extends to various quantum-inspired neural network architectures. 
}
\end{abstract} 

\maketitle

\section{Introduction}\label{sec:intro}

\noindent Quantum computing stands as a groundbreaking advancement that can tackle problems beyond the reach of classical computing, and thus revolutionizes fields like cryptography, material science, and optimization. Despite these advances, quantum noise, stemming from environmental interactions and hardware imperfections, poses a threat to the reliability and scalability of quantum computing~\cite{Preskill2018quantumcomputing,Arute2019supremacy,Morvan2024Phase,huang2020superconducting}. In response, Quantum Error Mitigation (QEM)~\cite{Endo2021Hybrid,Kandala2019Errormitigation,Li2017Efficient,Huang2023NearTerm,Cai2023Quantum} has become a critical field, offering qubit-conserving strategies to mitigate the noise impact on measurement results without the need for millions of qubits required by full quantum error correction~\cite{Google2024Quantum, Acharya2023Suppressing,bluvstein2024logical, Chiaverini2004Realization, Cory1998Experimental,Paetznick2024Demonstration,Reichardt2024Logical}, thus expediting near-term quantum computing applications. Standard QEM techniques, such as Zero-Noise Extrapolation (ZNE)~\cite{Kim2023Evidence,kim2023scalable,Endo2018Practical,Cai2021multi}, Probabilistic Error Cancellation (PEC)~\cite{vandenBerg2023Probabilistic,Song2019Quantumcomputation,Zhang2020Errormitigated}, and Clifford Data Regression (CDR)~\cite{Czarnik2021errormitigation,Strikis2021Learningbased,Piotr2022Improving}, can effectively reduce errors but face scalability issues due to their high sampling overheads~\cite{Quek2024exponentially,Takagi2022fundamental}, which scale up to exponentially with the circuit's gate number~\cite{Quek2024exponentially}. This exponential growth in sampling requirements poses significant hurdles for the practical application of QEM, particularly for complex quantum algorithms that involve deeper circuits. 

Within this landscape, Machine Learning for Quantum Error Mitigation (ML-QEM)~\cite{Liao2024Machine,Bennewitz2022Neural,liao2023flexible,Alexander2020adeeplearning,cantori2024synergy} stands out for its ability to retrieve noiseless results from noisy measurement data utilizing statistical model learning. In this manner, ML-QEM substantially cuts down on sampling overhead, thus economizing on quantum resources compared to standard QEM.
Recently, through comparative benchmarking of prevailing data-driven ML models~\cite{Liao2024Machine}—encompassing random forests (RF), multi-layer perceptrons (MLP), and graph neural networks (GNN)—across a spectrum of quantum tasks, ML-QEM has demonstrated significant outperformance over the standard QEM approaches with reduced sampling overheads. However, these conventional data-driven models, which are designed without specific physical context, empirically extract the intricate noisy-to-noiseless mappings. This learning pattern not only limits the physical interpretability of the models but also makes their efficacy highly dependent on the quantity and quality of the datasets. As quantum systems scale up, collecting data that meets such high standards becomes increasingly challenge, whether using real quantum devices or classical simulators. Notably, classical simulators are often relied upon to obtain the noiseless outputs of quantum circuits, which are then used as labels. Yet, the complexity of classical simulations required to generate noiseless outputs within quantum datasets grows exponentially with the system size, and this complexity can be further amplified, especially as the entanglement between quantum gates increases~\cite{Zhou2020WhatLimits,Kim2023Evidence}.

To address these challenges, we've developed a paradigm that deeply integrates QEM with artificial intelligence, taking a pioneering step towards tailoring machine learning models for QEM with physical priors. We introduce the Neural Noise Accumulation Surrogate (NNAS), a physics-inspired neural network designed for mitigating errors in complex quantum circuits. NNAS's modules are engineered to capture the cumulative effects of noise across quantum circuit layers, offering theoretical interpretability, and its alignment with physical processes is validated by experimental visualizations. This capability enables NNAS to emulate noise accumulation and predict its impact on measurements effectively. Our experiments demonstrate that NNAS maintains the strengths of ML-QEM, achieving state-of-the-art (SOTA) error mitigation with minimal quantum sampling overhead. It also overcomes the data-intensive nature of previous ML-QEM methods, training effectively with a limited dataset. Moreover, NNAS's proficiency in learning noise patterns enables effective error mitigation even in deeper circuits, a task typically difficult for QEM due to signal-to-noise challenges. Consequently, NNAS significantly enhances ML-QEM's capacity to handle complex quantum circuits, bolstering the utility of near-term quantum devices.
 
\section{Neural noise accumulation surrogate for error mitigation}
 
\begin{figure*}[t]
\includegraphics[width=1\textwidth]{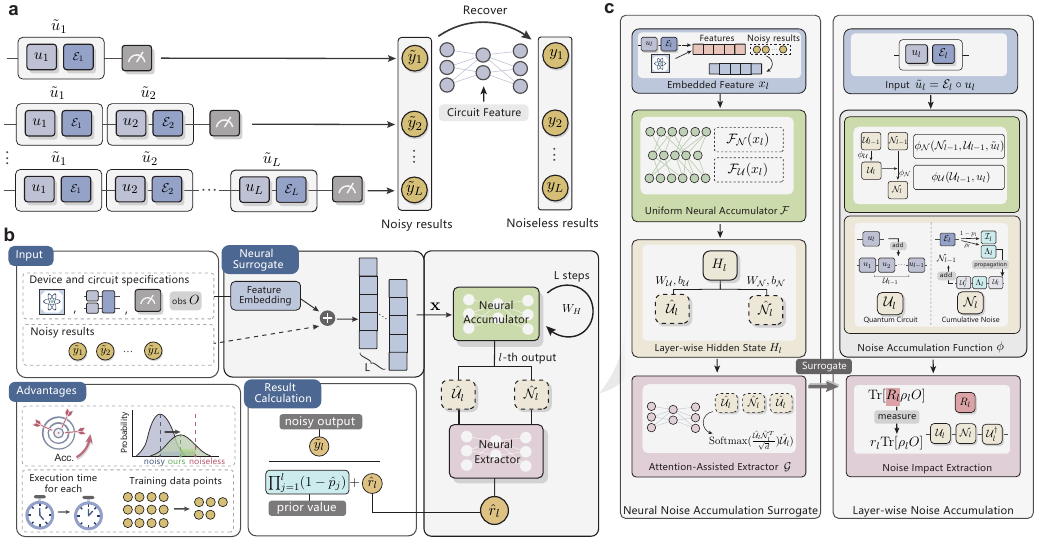}

\caption{\textbf{The architecture of the physical-inspired machine learning for quantum error mitigation.} \textbf{a} The illustration of layer-wise quantum sequential task. We model the quantum algorithms based on multi-layer circuits as quantum sequential tasks for a general analysis. The noisy quantum circuit comprises $L$ layers of gates $\tilde{\mathcal{U}}_L=\tilde{u}_L\circ...\circ \tilde{u}_2\circ\tilde{u}_1$, with each noisy layer $\tilde{u}_l$ decomposed into the noiseless layer $u_l$ and the noise channel $\mathcal{E}_l$, yielding the noisy results $\{\tilde{y}_l\}_l$ by measurement at each layer depth. We can recover the noiseless values $\{y_l\}_l$ from $\{\tilde{y}_l\}_l$ via machine learning (ML) for quantum error mitigation (QEM), with a neural network fed by descriptive features of circuits. \textbf{b} The overall pipeline of our Neural Noise Accumulation Surrogate (NNAS) for error mitigation. The input data including descriptive features for device and circuit specifications, and optional noisy result sequences are fed into NNAS, which encompasses three successive stages: embedding, neural accumulator, and neural extractor. The output of NNAS is the prediction of noise impact factors $\{\hat{r}_l\}_l$ for subsequent result calculation. The NNAS offers two key advantages: enhanced accuracy of mitigated values and reduced quantum resource requirements for training dataset. \textbf{c} Mapping relationship for constructing NNAS which conforms to the Layer-wise Noise Accumulation Framework designed for physical processes. The physical framework encompasses three components, which is mapped into NNAS as Embedded Feature, Uniform Neural Updater $\mathcal{F}$ along with the layer-wise hidden state $H_l$, and the Attention-Assisted Extractor $\mathcal{G}$. The embedded feature $x_l$ is obtained from the embedding of specification features, optionally incorporating noisy results. Within the Noise Accumulation Function tier of the physical framework, the cumulative noise $N_l$ at the $l$-th layer is updated based on $\mathcal{N}_{l-1}$. This update incorporates the twirled effective noise $\Lambda_l$ relocated to the end of the circuit $\mathcal{U}_l$ , where $\Lambda_l$ is derived from the layer-dependent noise channel $\mathcal{E}_l$ at $l$-th layer. In the Noise Impact Extraction, the noise envelope $R_l=\mathcal{U}_l\mathcal{N}_l\mathcal{U}_l^\dagger$ is quantified as the noise impact factor $r_l$ on the noiseless expectation value Tr$[\rho_lO]$ obtained from the noiseless output quantum state $\rho_l$ and observable $O$.}\label{fig1}
\end{figure*}

\noindent To facilitate a general analysis, we model the quantum algorithms based on multi-layer circuits as layer-wise quantum sequential tasks, as depicted in Fig.\ref{fig1} \textbf{a}. We examine the quantum circuit with $L$ layer of gates, denoted as $\mathcal{U}_L=u_L\circ u_{L-1}\circ ... u_1=\prod_{l=1}^Lu_l$, under the canonical Completely Positive and Trace-Preserving (CPTP)~\cite{nielsen00} noise postulate. 
The noisy quantum circuit, denoted as $\tilde{\mathcal{U}}_L$, can be expressed as the product of $L$ noisy layers: $\tilde{\mathcal{U}}_L=\prod_{l=1}^L\tilde{u}_l$. Each noisy layer is given by $\tilde{u}_l = \mathcal{E}_l \circ u_l$, where $\mathcal{E}_l$ represents the layer-dependent CPTP noise channel. 

In an effort to embed physical insights into the noisy-to-noiseless mapping that complements data-driven learning, we craft a model which harnesses layer-wise sequence data during training stage to capture the characteristics of noise accumulation patterns. Beyond mere sequence processing, our model is adaptable, capable of yielding output from a specific layer during prediction. This can be achieved by treating the output as a sequence and focusing on the target layer's result. Hence, we formulate the objective of the mitigation process by recovering the noiseless result sequences $\{y_l\}_{l=1}^L$ from noisy sequence $\{\tilde{y}_l\}_{l=1}^L$.

We then dissect the propagation of noise, aiming to isolate the noise impact to distill its cumulative effect within the circuit. By relocating each layer's noise channel to the end of circuit, we derive the noisy output state $\rho_l$ for the $l$-th layer:
\begin{equation}\label{eq-noisy_rho}
    \tilde{\rho}_l=(\prod_{j=1}^l(1-p_j))\rho_l+\left(\mathcal{U}_l \mathcal{N}_l \mathcal{U}_l^\dagger\right)\rho_l,
\end{equation}
with the effective rate $p_j$ and cumulative noise $\mathcal{N}_l$ derived from maximum noise decomposition (MND) of $\mathcal{E}_l$ (More details in \textquotesingle Methods\textquotesingle).
We define $R_l=\mathcal{U}_l \mathcal{N}_l\mathcal{U}_l^\dagger$ as the noise envelope and quantify its impact on results through the impact factor $r_l$.
Given an observable $O$, the task of recovering noiseless results is translated into the estimation of effective rates $\{\hat{p}_j\}_j$, which are the initial values obtained prior to mitigation (further details in Supplementary Section 1), and the impact factor $\hat{r}_l$ which is the output of our model. The mitigated expectation value is then calculated using the formula: 
\begin{equation}\label{eq-mev}
    y_l^{\text{em}}=\frac{\tilde{y}_l}{\prod_{j=1}^l(1-\hat{p}_j)+\hat{r}_l}.
\end{equation}
Isolating the noise envelope $R_l$ offers two significant advantages. First, it sharpens our focus on characterizing noise by directly assessing its impact on final results. Second, the estimated value $\prod_{j=1}^l(1-\hat{p}_j)$ can guide the trend of mitigated values across layers, aligning initial network training more closely with the target. This enhances the model's efficiency in terms of training resources and robust performance, particularly in deeper layers, as empirically validated in the subsequent {\textquotesingle Results\textquotesingle} section.

We refine our understanding of $R_l$ by developing a physical process framework that delineates its layer-by-layer dynamics, which serves as the foundational structure for our NNAS. This framework is segmented into three tiers, capturing the critical components of noise accumulation: input, noise accumulation function, and noise impact extraction. The NNAS is designed in accordance with this framework, aligning its modules with the framework's tiers, and it predicts the estimation of $r_l$ for the mitigated result calculation as given in Eq.~\ref{eq-mev}. The overall pipeline is depicted in Fig.\ref{fig1} \textbf{b}.

\begin{figure*}[htbp!]
\centering
\includegraphics[width=0.75\textwidth]{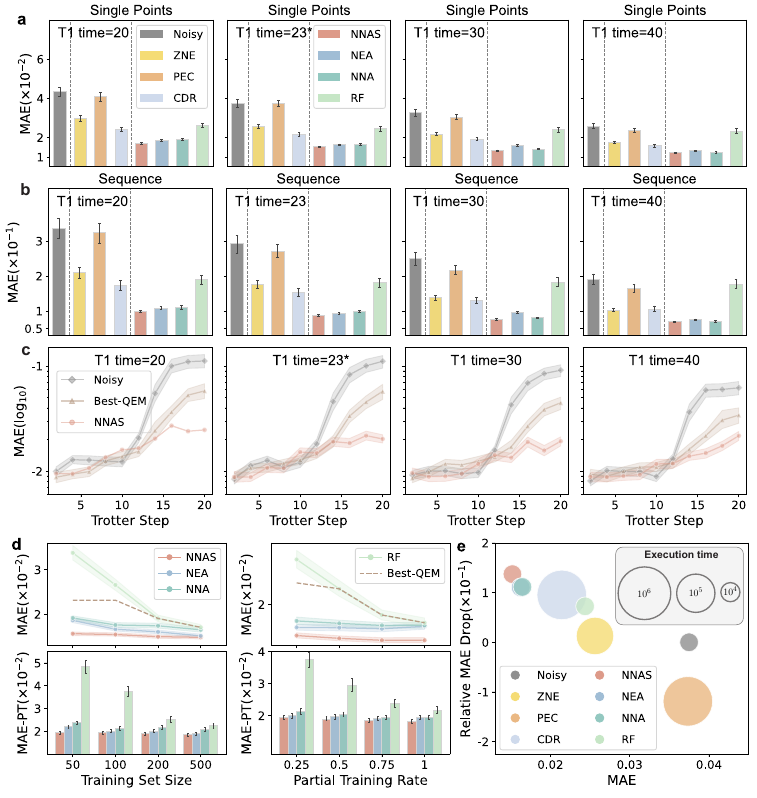}
\caption{\textbf{Evaluation results on Quantum Approximate Optimization Algorithm-type multi-layer circuits.} \textbf{a}-\textbf{b} The Mean Absolute Error (MAE) of baselines in different scenarios with increasing T1 time. The MAE is calculated using (upper) single points and (lower) sequence-derived vector norms. The Longer T1 time denotes lower noise level. We focus on T1 = 23.235 $\mu$s marked with 23*, which represent the state-of-the-art quantum coherence times~\cite{Morvan2024Phase}. \textbf{c} The comparison among Noisy, best-performing QEM methods including standard QEM methods and RF (denoted as Best-QEM), our model with growing Trotter step. We assess performance using single-point MAE computations. For \textbf{a}-\textbf{c}, we train the models using 100 sequences with partial training rate 0.25. \textbf{d} (upper) The MAE of full test set for all the ML-QEM methods with varying training set size and partial training rate. (lower) The MAE-PT, which represents of MAE for partial test set with layer depth 11-20, for ML-QEM methods. \textbf{e} The comparison of the sampling overhead which we quantify by execution time for mitigation during test stage among baselines, with marker size indicating the execution time. The performance metrics of our analysis are characterized by the Mean Absolute Error (MAE) and the drop in relative MAE, with the latter defined as the logarithmic ratio between the MAE of noisy and mitigated expectation values. Notably, a larger relative MAE drop signifies better performance, indicating a more effective mitigation of noise. The execution time required to mitigate a target quantum circuit, is given by the formula $T = N_c(C + N_s)$~\cite{Alireza2022Shadow} with some constant $C$. Following the precedent set in~\cite{Alireza2022Shadow}, we adopt $C=1000$ in our calculations. Here, $N_c$ denotes the total number of circuit instances needed to achieve a complete mitigation, and $N_s$ represents the number of measurement shots taken for each individual circuit instance. This formula is designed to capture the trade-off between altering the measurement basis for each circuit instance and the frequency of measurements within the same basis. All the data in the figure presents average results with 95\% confidence interval error bars.}\label{fig2}
\end{figure*}

\subsection{Uniform neural accumulator based on layer-wise hidden state}

\noindent The noise envelope $R_l$ consists of two components: the quantum circuit $\mathcal{U}_l$ and cumulative noise $\mathcal{N}_l$, forming the functional basis for noise accumulation. We reformulate them recursively in terms of the layer depth by
\begin{equation}\label{eq-accumu_U}
    \mathcal{U}_l=u_l\circ \mathcal{U}_{l-1}=\phi_{\mathcal{U}}(\mathcal{U}_{l-1},u_l),
\end{equation}
and
\begin{equation}\label{eq-accumu_noise}
    \begin{aligned}
        \mathcal{N}_l&=(1-p_l)\mathcal{N}_{l-1}+p_l\prod_{j=1}^{l-1}(1-p_j)a_l+p_la_l\circ\mathcal{N}_{l-1}\\
        &=\phi_{\mathcal{N}}(\mathcal{N}_{l-1},\mathcal{U}_{l-1},\tilde{u_l}),
    \end{aligned}
\end{equation}
respectively, to capture their progressive intricacies. Here $a_l=\mathcal{U}_l^\dagger \Lambda_l\mathcal{U}_l$ denotes the noise introduced by the $l$-th noisy layer and $\Lambda_l$ is the effective noise derived from MND of $\mathcal{E}_l$. We collectively term these recursive relations as the noise accumulation function $\phi$, synthesized from $\phi_{\mathcal{U}}$ and $\phi_{\mathcal{N}}$. The amalgamated input to $\phi$ is the noisy layer $\tilde{u}_l=\mathcal{E}_l\circ u_l$. 

To map these dynamics, we employ feature embedding on $M$ input variables, capturing the physical traits of the noisy layer and standardizing multi-structured quantum data into embedded data $X\in \mathbb{R}^{M*L}$, which serves as the input for RNN. To enhance model robustness, especially with limited descriptive data, we optionally concatenate the noisy results into $X$, with careful consideration and fine-tuning for specific tasks. 

Separate handling of the accumulation functions $\phi_{\mathcal{U}}$ and $\phi_{\mathcal{N}}$ results in loss of interwoven information. Despite the possibility of merging them at each step, this method leads to both error accumulation and prolonged total computation time. To address this, we propose a uniform neural accumulator $\mathcal{F}$, formulated as a recursive neural network (RNN)~\cite{Hochreiter1997Long,Schuster1997Bidirectional}, which integrates both $\mathcal{U}_l$ and $\mathcal{N}_l$ into a layer-wise hidden state $H_l$. This RNN is designed to capture the layer-depth-relevant dependence inherent in the evolution of both $\mathcal{U}_l$ and $\mathcal{N}_l$, ensuring a conformal mapping to the noise accumulation process that is both effective and computationally efficient. The hidden state $H_l$ is recursively calculated as follows:
\begin{equation}
\label{eq-RNN}
H_l = \mathcal{F}(X_l, H_{l-1}),
\end{equation}
where information accumulates progressively with layer depth. This recursive integration within $H_l$ continues until the neural extractor, where we distill the estimated noise impact factor $\hat{r}_l$. We employ learnable layer to recover the key components of $\mathcal{U}_l$ and $\mathcal{N}_l$ as $\hat{\mathcal{U}}_l$ and $\hat{\mathcal{N}}_l$ by
\begin{equation}
\begin{aligned}
    \hat{\mathcal{U}}_l &= W_{\mathcal{U}} \cdot H_l + b_{\mathcal{U}},\\
    \hat{\mathcal{N}}_l &= W_{\mathcal{N}} \cdot H_l + b_{\mathcal{N}},
\end{aligned}
\end{equation}
where $W_{\mathcal{U}}$, $W_{\mathcal{N}}$, $b_{\mathcal{U}}$ and $b_{\mathcal{N}}$ are learnable parameters.

\subsection{Attention-assisted noise impact extractor}

\noindent Aiming to capture the symmetric structure of the noise impact within $R_l = \mathcal{U}_l \mathcal{N}_l \mathcal{U}_l^\dagger $, we employ an attention mechanism tailored to align with this symmetry. Given that $\hat{\mathcal{U}}_l$ and $\hat{\mathcal{N}}_l$ are one-dimensional low-rank compressed vectors derived from $H_l$, directly emulating the matrix form $\mathcal{U}_l \mathcal{N}_l \mathcal{U}_l^\dagger$ is not feasible due to the dimensional constraints. Instead, we leverages an attention mechanism to merge information from the surrogates $\hat{\mathcal{U}}_l$ and $\hat{\mathcal{N}}_l$. The attention mechanism is formulated as follows:
\begin{equation}
    \text{A}_l = \text{Softmax}\left(\frac{\hat{\mathcal{U}}_l \hat{\mathcal{N}}_l^T}{\sqrt{d}}\right) \hat{\mathcal{U}}_l = \mathcal{G}(\hat{\mathcal{U}}_l, \hat{\mathcal{N}}_l),
\end{equation}
which introduces a computationally symmetric pattern that mirrors the conjugate symmetry of $R_l$.

The benefits of using the attention mechanism are twofold. First, the Softmax activation enhances the information density within $\hat{\mathcal{U}}_l \hat{\mathcal{N}}_l^T$ by normalizing attention scores, thereby highlighting the most pertinent information. Second, dividing by $\sqrt{d}$ enhances stability in the computation. This approach partially restores the compressed information and effectively fuses the representations from the low-rank compression of $\hat{\mathcal{U}}_l$ and $\hat{\mathcal{N}}_l$, achieving enhanced stability and efficacy by preserving the symmetrical structure with the attention-assisted extractor.
 
\section{Results}

\noindent We initially evaluate our model's performance on a typical class of multi-layer circuits, with exploration of the alignment of the structural features of the cumulative noise $\mathcal{N}$ and its surrogate $\hat{\mathcal{N}}_l$. Additionally, we validate NNAS's versatility by customizing it for a specific qubit-wise quantum algorithm task.
Baselines include the noisy case (Noisy), standard QEM methods (ZNE, PEC, CDR), typical ML-QEM methods using random forest (RF), and our ablated models: the neural extractor ablated (NEA) and the quantum-inspired neural network-both neural accumulator and extractor-ablated (NNA) (See Supplementary Section 2 for baseline details.). All baselines are trained with mean squared error (MSE).
We define the \textit{hard regime} in our dataset where sequences exceed half the maximum sequence length, indicating higher complexity in data acquisition. And we denote the proportion of this regime in the training set by the partial training rate $p_r$. The detailed experimental settings and evaluation metrics are given in Supplementary Section 3. 

\begin{figure*}[t]
\centering
\includegraphics[width=0.93\textwidth]{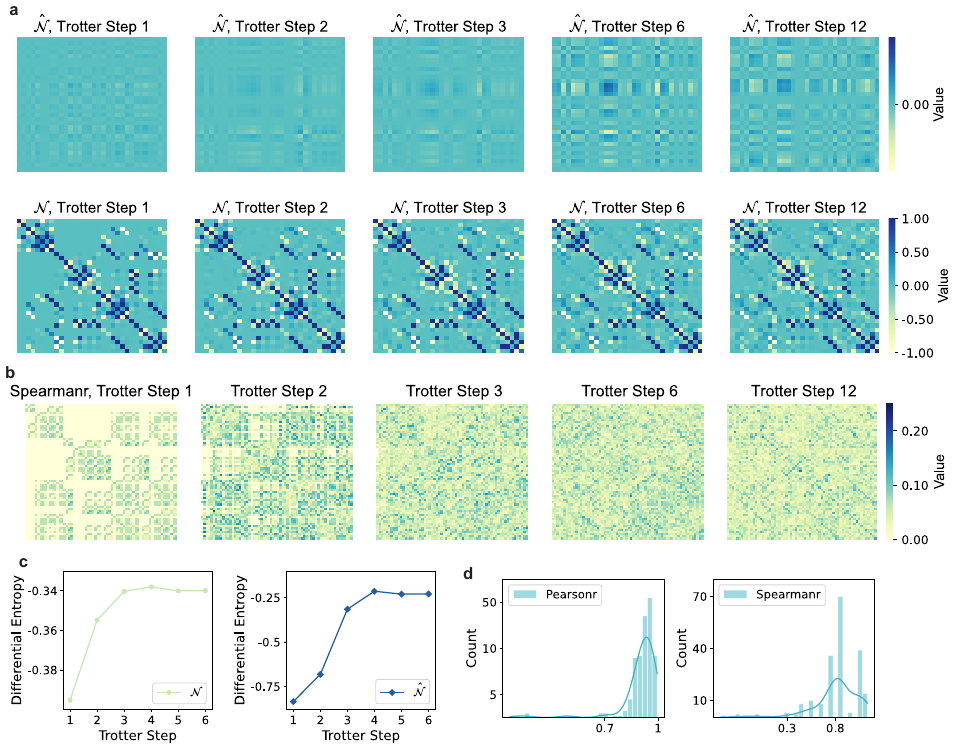}
\caption{\textbf{Comparative visualization and statistical analysis for the structural correspondence between quantum cumulative noise and its surrogate.} \textbf{a} The visualization of the matrix of surrogate $\hat{\mathcal{N}}^T \hat{\mathcal{N}}$ (upper) and the representation of the cumulative noise $\mathcal{N}$ using the compressed Pauli Transform Matrix~\cite{Greenbaum2015Introduction} (PTM) with sum pooling (lower). \textbf{b} The visualization of the Spearman correlation matrix illustrates the relationships between the surrogate $\hat{\mathcal{N}}$ and the compressed PTM of $\mathcal{N}$. The values of each matrix are computed using the vectors of $\hat{\mathcal{N}}$ and the corresponding windows in the compressed PTM of $\mathcal{N}$ that match the shape of $\hat{\mathcal{N}}^T \hat{\mathcal{N}}$ . \textbf{c} The curve of differential entropy for the surrogate $\hat{\mathcal{N}}$ and the compressed PTM of $\mathcal{N}$ with growing Trotter steps. Experiments \textbf{a}-\textbf{c} constitute a case study on a single sample instance. The observed increase in disorder, mirrored by the corresponding trend in $\hat{\mathcal{N}}$, indicates the escalating effect of cumulative noise on system dynamics, thereby demonstrating the effectiveness of our model in capturing these structural characteristics. \textbf{d} The probabilistic distribution of the Pearson (left) and Spearman (right) correlation coefficients between the entropy curves of $\mathcal{N}$ and its surrogate $\hat{\mathcal{N}}$, evaluated across 200 sequences in the test set.}\label{fig3}
\end{figure*}
 
\subsection{Evaluations on typical layer-wise quantum circuits---QAOA-type circuits}

\noindent Our evaluations are centered around Quantum Approximate Optimization Algorithm (QAOA)-type circuits, which are crucial for near-term quantum algorithms, including quantum Trotterized simulations~\cite{Smith2019Simulating,hu2023symmetric,rendon2024improved} and variational quantum algorithms (VQAs)~\cite{cao2019quantum,McArdle2020Quantum,Bharti2021NoisyISQ}. These QAOA-type circuits are distinguished by their layer units, which are determined by a certain Hamiltonian. In our experiments, we focus on the temporal evolution of the 1D transverse-field Ising spin chain~\cite{Smith2019Simulating}, as described by the Hamiltonian $H_\text{Ising}(J,h)$ with circuit parameters $J$ and $h$. We utilize the first-order Trotterized circuit, a standard configuration for QAOA-type circuits in quantum Trotterized simulations. All machine learning models are trained on sequence data, and the Mean Absolute Error (MAE) is calculated using both sequence norms and single point methods. This allows us to validate the effectiveness of our model in two applications that leverage QAOA-circuits: quantum Trotterized simulations and VQAs, respectively. We evaluate our model's performance on six-qubit circuits with random circuit parameters, varying the layer depth (Trotter step) from 1 to 20. Our training set encompasses a hard regime including on steps 11-20. In the testing stage, we assess all baselines across 200 sequences, spanning the full range of Trotter steps from 1 to 20.
 
\noindent\textbf{Result Comparisons.}
Our model demonstrates superior performance over baseline models across various noise levels, particularly excelling in high-noise scenarios as shown in Fig.\ref{fig2} \textbf{a} and \textbf{b}. Notably, this achievement comes with training on a limited dataset of 100 sequences at a partial training rate of $p_r=0.25$, highlighting the efficiency of our model in data usage. Our model significantly reduces MAE by an average of 65.85\% across two key calculations compared to the noisy case (Noisy) and by 35.12\% over the best-performing QEM methods, including standard QEM and RF approaches. This advantage is accentuated at larger Trotter steps, especially beyond Trotter step 15, where our model achieves a 79.42\% MAE reduction compared to Noisy and a 55.63\% improvement over the best-performing QEM methods as depicted in Fig.\ref{fig2} \textbf{c}. These results underscore the scalability and robustness of our model's theoretical design.

We then explore our model's efficiency compared to baselines using ML models, including RF and our ablated models, under certain noise conditions that reflect SOTA quantum computational capabilities~\cite{Morvan2024Phase}. As given in Fig.\ref{fig2} \textbf{d}-\textbf{e}, the former focuses on the training phase, while the latter focuses on the prediction phase.
In Fig.\ref{fig2} \textbf{d}, while RF exhibits a pronounced sensitivity to changes in dataset scale, NNAS demonstrates robustness to this variations. This is evidenced by its ability to maintain accuracy with a 90\% reduction in dataset size and a 75\% decrease in $p_r$, along with an order-of-magnitude reduction in the total number of data points across all sequence data. We proceed to demonstrate the sampling overhead for mitigation, which we quantify by the total execution time required to mitigate a target quantum circuit (See detailed settings in the Supplementary Section 2), among baselines during prediction, as depicted in Fig.\ref{fig2} \textbf{e}. Our model achieves peak accuracy with the sampling overhead reduced by 1-2 orders of magnitude compared to standard QEM methods, emphasizing its efficiency and practicality.

\noindent\textbf{Explorations on how neural networks surrogate the cumulative noise.}
The comparative results above, including results of ablated models NEA and NNA with an average MAE reduction of 8.58\% over NEA and 6.99\% over NNA, preliminarily confirm that the synergistic integration of the nueral noise accumulator and extractor into our model significantly enhances performance. Building on these findings, we will further substantiate our model's enhanced performance by delving into its underlying mechanism. We aim to investigate the structural alignment between the quantum cumulative noise $\mathcal{N}$ and its surrogate $\hat{\mathcal{N}}$, which is crucial for assessing the model's ability to embed structural information. Our analysis begins with a case study, as given in Fig.\ref{fig3} \textbf{a}-\textbf{c}.
Fig.~\ref{fig3} \textbf{a} presents the structural evolution of $\mathcal{N}$ and $\hat{\mathcal{N}}$, visualized using the Pauli Transfer matrix~\cite{Greenbaum2015Introduction} (PTM) for $\mathcal{N}$ with compression via sum pooling and the two-dimensional matrix $\hat{\mathcal{N}}^T \hat{\mathcal{N}}$ for $\hat{\mathcal{N}}$. With increasing Trotter steps, the distribution across these matrices shows increased dispersion, indicating a progressive alignment of structural changes between $\mathcal{N}$ and $\hat{\mathcal{N}}$.
Fig.~\ref{fig3} \textbf{b} extends the analysis by computing the Spearman rank correlation matrix between $\hat{\mathcal{N}}^T \hat{\mathcal{N}}$ and sub-matrices of $\mathcal{N}$. The initial block diagonal-like pattern in the correlation matrix, due to sparse off-diagonal elements of $\mathcal{N}$, transitions to a rapid divergence in correlation values from steps 1 to 3, highlighting the model's swift adaptation to early structural changes. This is followed by a decrease in local correlation, emphasizing the model's focus on global characteristics of $\mathcal{N}$ over specific details, especially as dispersion gradually increases in later steps.
Building on this, Fig.\ref{fig3} \textbf{c} demonstrates the alignment of differential entropy curves between $\mathcal{N}$ and $\hat{\mathcal{N}}$ across increasing Trotter steps, using differential entropy to gauge the diversity within the noise profiles.
Expanding on these insights, Fig.\ref{fig3} \textbf{d} conducts a statistical evaluation across 200 test sequences, affirming the model's consistent replication of quantum noise's structural dynamics. High average correlation coefficients—Pearson at 0.8991 and Spearman at 0.8028—calculated from the entropy curves of $\mathcal{N}$ and $\hat{\mathcal{N}}$, substantiate the broad validity of our findings. Collectively, these analyses confirm the model's proficiency in capturing the structural nuances of the physical process.

\begin{figure*}[t]
\centering
\includegraphics[width=0.8\textwidth]{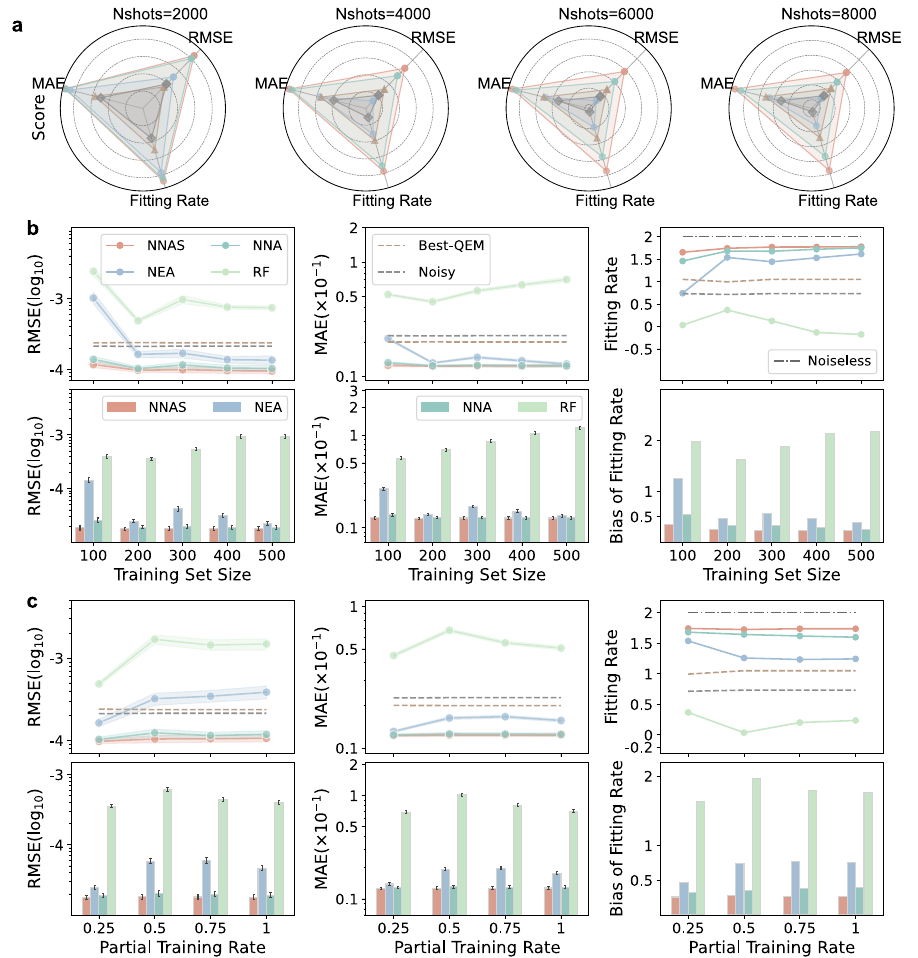}
\caption{\textbf{Evaluation results on quantum metrology (QM) based on GHZ probe state.} QM focuses on the enhancement of measurement sensitivity as the qubit number scales up. We employ three critical metrics for QM, including the root mean square error (RMSE) for the estimated parameter $\theta$, the MAE for the mitigated expectation value, and the fitting rate of the RMSE scaling curve with respect to qubit number. The RMSE and fitting rate serve as indicators for QM, with the latter being an upper-bounded metric where values closer to 2 signify better performance, as they approach the noiseless scenario. The MAE, alongside the RMSE, quantifies the error in our estimates. \textbf{a} The score of three metrics to evaluate the performance on QM. For RMSE and MAE, the score represents the ratio of the noiseless value to the mitigated value, given a finite number of measurement shots. Conversely, for the fitting rate, the score is the ratio of the mitigated value to the noiseless value. We train the ML models on training set with 200 sequences, including GHZ states of 1-10 qubits with hard regime including number of qubits 6-10 and partial training rate 0.25. \textbf{b}-\textbf{c}, (upper) The metrics of full test set for ML-QEM methods with varying training set size and partial training rate. (lower) The metrics of partial test set with qubit number 6-10 (for RMSE and MAE), and the bias of fitting rate between the mitigated value and noiseless value. All the data in the figure presents average results with 95\% confidence interval error bars.}\label{fig4}
\end{figure*}

\subsection{Extending to qubit-wise quantum circuits}

\noindent In this extended experiment, we showcase NNAS's broad applicability by adapting it to specific qubit-wise quantum algorithms, which focus on their performance enhancement as the qubit number scales up. NNAS can be adapted to scenarios where the configuration of initial qubits remains unchanged as additional qubits are integrated. In such cases, we can regard the growth in the number of qubits as equivalent to the growth in layers, treating the maximum qubit count as the upper limit for layer-wise expansion within the existing model framework. To exemplify this extended application, we turn to quantum metrology (QM)~\cite{giovannetti2011advances}, a domain where the precision of quantum measurements is directly linked to the number of qubits. We employ the Greenberger-Horne-Zeilinger (GHZ) entangled state as the probe state for QM. The circuit for preparing an $n$-qubit GHZ state is expressed as follows:
\begin{equation}
    U^{\text{GHZ}}_n = \prod_{i=1}^{n-1} \text{CNOT}_{i,i+1} = U^{\text{GHZ}}_{n-1} \cdot \text{CNOT}_{n-1,n},
\end{equation}
where $\text{CNOT}_{i,i+1}$ denotes the CNOT gate with the $i$-th qubit as control and the $(i+1)$-th qubit as target. We leverage the structural equivalence of the GHZ state preparation circuit, where each CNOT gate corresponds to a layer in the circuit, directly relating the number of layers to the number of qubits. For each circuit parameter $\theta$, we evolve the GHZ states $\{\text{GHZ}_n\}_n$ under the unitary operations $\{U(\theta)_n\}_n$ and measure them with the global Pauli $X$ observables $\{X^{\otimes n}\}_n$. Our dataset includes GHZ states with 1 to 10 qubits, focusing on the hard regime of qubits 6-10. Model training involves sequences parameterized by circuit parameters randomly sampled between 0.1 and 5.0 with 0.1 increments, with 10 independent trials per parameter. Evaluations use parameters from 0.05 to 5.05 in 0.1 steps.

To evaluate our model's performance, we employ three critical metrics for QM, including the root mean square error (RMSE) for the estimated parameter $\theta$, the MAE for the mitigated expectation value, and the fitting rate of the RMSE scaling curve with respect to qubit number. The RMSE and fitting rate serve as indicators for QM, with the latter being an upper-bounded metric where values closer to 2 signify better performance, as they approach the noiseless scenario. The MAE, alongside the RMSE, quantifies the error in our estimates. A comparative analysis of these metrics across various measurement shot numbers $N_s$ for baselines is presented in Fig.~\ref{fig3} \textbf{a}.
Despite trained on a limited dataset of 200 sequences derived from just 20 distinct angles with $p_r=0.25$, our model significantly outperforms Noisy by 52.81\% to 84.95\% and the best-performing QEM methods by 46.92\% to 53.43\% across three key evaluation metrics.
Figs.~\ref{fig3} \textbf{b} and \textbf{c} demonstrate that our model achieves precision advantages with a limited dataset scale compared to other baselines using ML models. Notably, our model realizes a nearly two orders of magnitude reduction in RMSE and a one order of magnitude reduction in MSE relative to the RF model, underscoring the superior efficiency of NNAS in the extended application for mitigating qubit-wise quantum algorithms.

\section{Conclusion}

\noindent Our proposed NNAS demonstrates SOTA performance in results’ error mitigation, quantum sampling overhead, and scalability of training datasets, significantly outperforming previous QEM methods. This advancement is attributed to our approach, which goes beyond the conventional ML-QEM by not just employing established machine learning models for error mitigation, but by harmoniously integrating the structural characteristics of quantum noise accumulation within the neural network's architecture. This integration endows our network with physical interpretability and reduces the reliance on the size and scope of quantum data sets. It's important to note that our focus on multi-layer circuits is universally applicable in quantum computing, meaning our NNAS is nearly adaptable to all current quantum circuits. This significantly enhances the capabilities of QEM, enabling near-term quantum devices to handle larger-scale and deeper circuits, thus paving the way for practical applications.

Moreover, our contribution transcends the realm of QEM; it represents a paradigm that integrates neural network architecture with the evolution of physical states. Our framework for physical processes encapsulates quantum noise accumulation across three dimensions: input, state transition, and output. This multi-tiered structure not only accommodates quantum noise accumulation but also aligns with the overarching principles of physical state evolution, underscoring its adaptability as a general paradigm for physical processes that extend beyond the purview of QEM. Our study highlights the potential of tailoring ML network structures for specific quantum tasks, thereby paving the way for future research in quantum technologies that can achieve high performance with limited quantum datasets.

\section{Methods}
 
\renewcommand{\thefigure}{E\arabic{figure}}
\setcounter{figure}{0}
\begin{figure*}[t]
\centering
\includegraphics[width=1\textwidth]{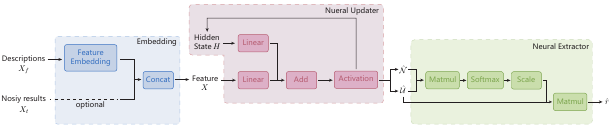}
\caption[Extended Data Fig. 1]{\textbf{The architecture of NNAS}. The architecture mainly consists of three parts. (1) Embedding: We realize the learnable embedding methods to extract relevant information from input and concatenate them together as the feature $X$. (2) Neural Updater: The recurrent network with hidden state implicitly represent the compressed cricuit and noise matrix, simulates the recursive noise accumulation process. (3) Neural Extractor: The modified attention mechanism effectively surrogate the noisy term. }\label{fig5}
\end{figure*}

\subsection{Layer-wise quantum noise accumulation}

\noindent The physical framework for layer-wise quantum noise accumulation is established on the basis of the maximum decomposition of layer-dependent noise and the formulation of noise accumulation. We will now detail their derivation as follows.

\noindent\textbf{Maximum noise decomposition.}
For an $n$-qubit layer-dependent CPTP noise channel $\mathcal{E}_l$ for noisy layer $\tilde{u}_l$, we define the maximum noise decomposition (MND),
\begin{equation}\label{eq-MND}
    \mathcal{E}_l=(1-p_l)\mathcal{I}+p_l\Lambda_l,
\end{equation} 
with $\Lambda_l$ representing the effective noise channel, and $p_l$ being the effectiveness factor satisfying
\begin{equation}\label{eq-define_p}
    \min\left(\lambda\{C_{[\mathcal{E}_l-(1-p_l)\mathcal{I}]}\}\right)=0, \ 0\leq p_l\leq 1.
\end{equation}
We employ the PTM for detailed analytical deductions. PTM is a useful representation for any $n$-qubit quantum channel $A$:
\begin{equation}
    {(\mathcal{R}_\mathcal{A})}_{ij}=\frac{1}{2^n}\text{Tr}[P_i\mathcal{A}(P_j)],    
\end{equation}
where $P_i$ is the $i$-th Pauli operator from Pauli group $\mathcal{P}^{\otimes n}$ where $\mathcal{P}=\{I,X,Y,Z\}$. In the PTM representation, the CPTP channel's essential properties of complete positivity (CP) and trace preservation (TP) are reflected as:
\begin{itemize}
\item CP constraint: The Choi-Jamiolkowski (CJ)~\cite{nielsen00,homa2024choi} matrix $\mathcal{C}_\mathcal{A}$, which can be written in terms of the PTM as  
\begin{equation}
    \mathcal{C}_\mathcal{A}=\frac{1}{2^{2n}}\sum_{i,j=1}^{d^2}(\mathcal{R}_\mathcal{A})_{i,j}P_j^T\otimes P_i,
\end{equation}
satisfys $\mathcal{C}_\mathcal{A}\geq 0$.
\item TP constraint: ${(\mathcal{R}_\mathcal{A})}_{0j}=\delta_{0j}$.
\end{itemize}
The main purpose of the MND is to isolate the maximal identity component within the noise channel, enabling our model to more accurately characterize the noise behavior. During decomposition defined in Eq.~\ref{eq-MND}, we constrain the values of $p_l$ to ensure that $\Lambda_l$ remains a CPTP channel, thus guiding the model to adhere to physical laws. Since the TP constraint $(\mathcal{R}_{(\mathcal{E}_l-(1-p)\mathcal{I}})_{0j}=\delta_{0j}$ holds for any $0\leq p_l\leq 1$, the $p_l$ is defined as in Eq.~\ref{eq-define_p} to satisfy the CP constraint.

We illustrate the MND with the typical incoherent Pauli noise channel. 
For layer $l$ and error rate $p$, the $n$-qubit Pauli noise channel is
\begin{equation}
    \mathcal{E}^l_p=\sum_{i=0}^{4^n-1}\delta_{i,p}^l\mathcal{P}_i(\cdot),
\end{equation}
where $\mathcal{P}_i$ is the super-operator of the $i$-th Pauli operator and $\{\delta_{i,p}^l\}_i$ are the coefficients of Pauli terms which satisfy $\delta_{i,p}^l\geq 0$, $\sum_i\delta_{i,p}^l=1$. We can easily calculate the effectiveness factor of $\mathcal{E}^l_p$ that $p_l=\sum_{i>0}\delta_{i,p}^l=1-\delta_{0,p}^l$, which is the total error probability of the non-trivial Pauli terms.
For the incoherent Pauli noise channel, the effectiveness factor $p_l$ is correlated with the fidelity decay rate, which can be efficiently estimated using benchmarking protocols ~\cite{magesan2012efficient,emerson2007symmetrized,Helsen2022General}. However, in the case of coherent noise channels, where noise is congruent with unitary operations, the effectiveness factor $p_l$ is unity, rendering that the MND doesn't work. Yet our model can still be employed to learn the effectiveness of the noise channel $\mathcal{E}_l$. Furthermore, by employing noise tailoring strategies, such as randomized compiling ~\cite{Wallman2016Noisetailoring,Erhard2019Characterizing,Hashim2021Randomizedcompiling}, we can transform any CPTP noise channel into an incoherent Pauli noise channel.

\noindent\textbf{Cumulative noise update function.}
By using MND, the noisy quantum circuit $\tilde{\mathcal{U}_L}$ with $L$ layers,
\begin{equation}\label{eq-recurrentU}
    \tilde{\mathcal{U}}_L=\prod_{l=1}^L \mathcal{E}^l\circ u_l=\prod_{l=1}^L [(1-p_l)\mathcal{I}+p_l\Lambda_l]\circ u_l, 
\end{equation}
can be expressed in a form analogous to the binomial theorem expansion. After merging terms, it reads
\begin{equation}\label{eq-noisyU}
    \tilde{\mathcal{U}}_L=\prod_{l=1}^L [(1-p_l)\mathcal{U}_L+[\mathcal{U}_L\mathcal{N}_L\mathcal{U}_L^\dagger] \mathcal{U}_L,
\end{equation}
with the cumulative noise $A_L$ gives, 
\begin{equation}\label{eq-recurrentN}
    \mathcal{N}_L=\sum_{M=1}^L \sum_{k=1}^{\binom{L}{M}}(\prod_{j\in s_M^k}p_j\mathcal{U}_j^\dagger\Lambda_j\mathcal{U}_j)\cdot\prod_{j'\notin s_M^k}(1-p_j'),
\end{equation}
Here $s_M^k$ represents the set of indices for the $k$-th occurrence of effective noise across $M$ layers.

The iterative update formula of the cumulative noise $\mathcal{N}_l$ in terms of the layer number $l$ is given in Eq.~\ref{eq-accumu_noise}, which can be intuitively understood as follows. After introducing the $l$-th noisy layer $\tilde{u}_l$, the event of the occurrence of effective noise in the circuit $\tilde{\mathcal{U}}_l$ can be delineated into three discrete scenarios:
\begin{itemize}
    \item The $\tilde{u}_l$ does not introduce effective error, denoted as $(1-p_l)\mathcal{N}_{l-1}$.
    \item The $\tilde{u}_l$ introduces effective error while the preceding $l-1$ layers remain error-free, characterized as $p_l\prod_{j=1}^{l-1}(1-p_j)a_l$.
    \item The $\tilde{u}_l$ introduces effective error along with errors occurring in the preceding $l-1$ layers, denoted as $p_l a_l \mathcal{N}_{l-1}$.
\end{itemize}  

\subsection{Detailed construction of physics-inspired neural network}


\noindent We will next detail the computational procedure of the NNAS model to understand its approach to modeling the noise accumulation process within the multi-layer circuits. The procedure is broadly divided into three key steps:

\begin{itemize}

\item Step 1: The uniform neural accumulator recurrently simulate the noise accumulation process.

\item Step 2: Recover the layer-wise implicit representation $\mathcal{N}_l$ and $\mathcal{U}_l$ from hidden state $H_l$ in each layer $l$. 

\item Step 3: The Attention-assisted extractor combine the information of $\mathcal{N}_l$ and $\mathcal{U}_l$ and calculate $r_l$ with a readout linear layer.

\end{itemize}
The architecture of NNAS is illustrated in Extended Data Fig.\ref{fig5} the comprehensive algorithm is outlined in Alg~\ref{alg-overall}. 

\begin{algorithm}[H]
\label{alg-overall}
\caption{Neural Noise Accumulation Surrogate}
\SetKwFunction{FMain}{}
\SetKwProg{Fn}{Function}{:}{}
\Fn{\FMain{$G,N$}}{
    $G$ represents the device and circuit specifications, $N$ represents the noisy results, $P$ represents the pre-processed effectiveness factors
    
    $X_f \gets \text{Feature\_Embed}(G)$
    
    $X \gets \text{concat}([X_f, N])$

    \For{$l \gets 1 \ to \ L$}{
        $H_l \gets \text{RNN\_Cell}(X_l)$ \Comment{Step 1}
        
        $\text{Recover} \ \mathcal{U}_l, \ \mathcal{N}_l\ \text{from} \ H_l$ \Comment{Step 2}
        
        $A_l \gets \text{Softmax}(\frac{\mathcal{U}_l \cdot \mathcal{N}_l^T}  {\sqrt{d}}) \mathcal{U}_l$ \Comment{Step 3}
        
        $r_l \gets \text{Linear\_Readout}(A_l)$

    }
    $ y \gets \frac{\mathcal{N}} { (P + r)} + b$ \Comment{b is a series of parameter to learn the overall deviation}
    
    \textbf{return} $y$ \Comment{The prediction results}
    }
\end{algorithm}

\noindent\textbf{Feature embedding.}
The input data typically include specific information of the device and circuit and the observed noisy results. We delve into the embedding layers to extract relevant information from original input data, which aims to enhance the model's ability for recognizing different circuit instances.

We use $M$ variables to describe the specifications of device and circuit, such as quantum system size, type of the circuit, type of the noise, error rate for single gates, error rate for double gates, and circuit parameters. The realization of the feature embedding diverse according to the form of the specification data. 

\begin{itemize}
    \item For single discrete values, formally $G\in \mathbb{N}^{1*1}$, such as quantum number and circuit type, we use a learnable embedding function $E  = \mathrm{f(G)} = G W_1$ to embed it, where $W_1 \in \mathbb{R}^{1*1}$ is a learnable scale factor to scale the discrete values into appropriate range. Then we encoded it with linear layer $X^{sd}=\mathrm{g(E)} = E W_2 + b$ where $W_2 \in \mathbb{R}^{1*L}$ and $b \in \mathbb{R}^{1*L}$ are learnable parameters, and we obtain the embedded feature $ X^{sd}\in \mathbb{R}^{1*L}$. 
    \item  For single continuous values, formally $G \in \mathbb{R}^{1*1}$, such as delta, we use a linear function $\mathrm{E = f(G)} = G W_1 + b$ to embed it, where $W_1 \in \mathbb{R}^{1*1}$ is a learnable scale factor to scale the discrete values into appropriate range and $b \in \mathbb{R}^{1*1}$ is a learnable offset. Then we encoded it with linear layer $X^{sc}=\mathrm{g(E)} = E W_2 + b$ where $W_2 \in \mathbb{R}^{1*L}$ and $b \in \mathbb{R}^{1*L}$ are learnable parameters, and we obtain the embedded feature $ X^{sc}\in \mathbb{R}^{1*L}$. 
    \item For multi-variables with discrete values, formally $G \in \mathbb{N}^{1*s}$, such as observables. Defining $s$ is the size of variables, we transform it into  $G^\text{T} \in \mathbb{N}^{s*1}$ and fit it with an embedding function $E  = \mathrm{f(G)} = G W_1$, where $W_1 \in \mathbb{R}^{s*L}$ is a set of learnable parameters to decompose the impact of $G$ to each circuit layer. Then we use a linear layer to combine the impact of all the variables, namely $X^{md}=\mathrm{g(E)} = E^T W_2 + b$, where $W_2 \in \mathbb{R}^{s*1}$ and $b \in \mathbb{R}^{1*1}$ are learnable parameters. Finally, we transpose it and obtain the embedded feature $ X^{md}\in \mathbb{R}^{1*L}$.  
\end{itemize}

Besides, to enhance model robustness, especially with limited descriptive data, we optionally integrate noisy results $X^{n}\in \mathbb{R}^{1*L}$ with embedded descriptions to extract latent statistical patterns.

All kinds of specifications with different forms are properly embedded with suitable methods to map the global information to each layer. Then they are concatenated together as $X^f \in \mathbb{R}^{M*L}$, where $M$ is the total kind of specific variables and $L$ is the length of circuit.

\noindent\textbf{Uniform neural accumulator simulating the noise accumulation.}
As $\mathcal{U}_l$ and $\mathcal{N}_l$ are recursively cumulated in equations \ref{eq-recurrentU} and \ref{eq-recurrentN}, we naturally opt for an RNN to simulate their iteration process. Formally, the recurrent process can be expressed as 
\begin{equation}
\begin{aligned}
   &\mathcal{N}_l=\phi_\mathcal{N}(X_{l},\mathcal{N}_{l-1},\mathcal{U}_{l-1}), \\
   &\mathcal{U}_l=\phi_\mathcal{U}(X_{l},\mathcal{N}_{l-1},\mathcal{U}_{l-1}). 
\end{aligned}
\end{equation}
With $\mathcal{N}_l$ and $\mathcal{U}_l$ satisfying the cross-spread relation, we utilize a hidden state $H_l$ to implicitly parameterize them. 
And the recurrent function $\phi_\mathcal{N}(\cdot)$ and $\phi_\mathcal{U}(\cdot)$ can be implicitly expressed through the recurrent neural network function $\mathcal{F}$, which is formulated as:
\begin{equation}
 H_l=\mathcal{F}(X_l, H_{l-1}) = \text{RNN}(X_l, H_{l-1}),
\end{equation}

The RNN cell consists of two linear projections and an activate operator. Specifically, in each iteration, the hidden state $H_l$ of the circuit layer $l$ is calculated as:
\begin{equation}
 H_l=\tanh(\text{Linear}_x(X_l) + \text{Linear}_h(H_{l-1})),
\end{equation}
where $\tanh$ is an activate operator and the projection $\text{Linear}(\cdot)$ with learnable parameter $W$ and $b$ can be formulated as:
\begin{equation}\label{eq-linear}
\text{Linear}(X) = W\cdot X + b.
\end{equation}

The hidden state $H_l$ implicitly represents the compressed cross-spread pattern of $\mathcal{N}_l$ and $\mathcal{U}_l$. The hidden state $H_l$ will not be decomposed in the recurrent process to avoid introducing additional accuracy errors and reducing efficiency.

\noindent\textbf{Attention-assisted extractor based on a computationally similar attention core.}
The Attention mechanism~\cite{Attention} has revealed a promising dependencies alignment ability in sequence learning~\cite{Informer, Transformer4Seq2, Transformer4Seq3}, which can be defined as:
\begin{equation}\label{eq-valina-attn}
    \mathrm{Attn}(Q, K, V)=\mathrm{Softmax}(\frac{Q\cdot K^T}{\sqrt{d}})V,
\end{equation} 
where $d$ is the scaled factor of data dimension, and $Q$, $K$ and $V$ are linear projected from input data $X$. 

Though observing the mathematical formulation of noise envelope $R_l=\mathcal{U}_l\mathcal{N}_l\mathcal{U}_l^\dagger$, we introduce the attention mechanism as the neural extractor to surrogate the calculation of $R_l$. Then, we represent the hidden components of $\hat{\mathcal{N}}_l$ and $\hat{\mathcal{U}}_l$ from $H_l$ in the following way:
\begin{equation} \label{eq-H-Ul}
\begin{aligned}
    \hat{\mathcal{U}}_l &= W_{\mathcal{U}} \cdot H_l + b_{\mathcal{U}},\\
    \hat{\mathcal{N}}_l &= W_{\mathcal{N}} \cdot H_l + b_{\mathcal{N}}.
\end{aligned}
\end{equation}
Thus, we substitute the new definition into vanilla attention mechanism.
\begin{equation}
    A_l=\mathrm{Softmax}(~(\frac{\hat{\mathcal{U}}_l\hat{\mathcal{N}}_l^\text{T}}{\sqrt{d}})\hat{\mathcal{U}}_l~),
\end{equation} 
where the hidden states $H_l$ are transformed into $\hat{\mathcal{N}}_l$ and $\hat{\mathcal{U}}_l$ that simulate the ${\mathcal{N}}_l$ and ${\mathcal{U}}_l$, respectively. After that, the linear projection is defined as a simple readout layer on $A_l$.

\section*{Acknowledgements}
This work was supported by the National Science and Technology Major Project (No. 2022ZD0117800). H.-L. H. acknowledges support from the National Natural Science Foundation of China (Grant No. 12274464), and Natural Science Foundation of Henan (Grant No. 242300421049).

\section*{Declarations}

\subsection*{Author contribution}
H.-L. H. initiated the project. H.-L. H., H.-Y. Z. crafted the general workflow for this study. H.-L. H, H.-Y. Z. and W.-S. B. supervised the work. X.-Y. X., X. X. and T.-Y. C. proposed the framework of the Neural Noise Accumulation Surrogate. X.-Y. X. and X. X. conducted the experiments and analyzed the results. X.-Y. X., X. X., H.-L. H. and H.-Y. Z. wrote the paper. All authors contributed to the discussion of the results.

\subsection*{Competing interests}
The authors declare no competing interests.

\bibliography{b}

\end{document}


\title{Supplementary information: Physics-inspired Machine Learning for Quantum Error Mitigation}
\maketitle

\section{The estimation of effectiveness factors}
\begin{figure}[t]
    \centering
    {\includegraphics[width=0.9\textwidth]{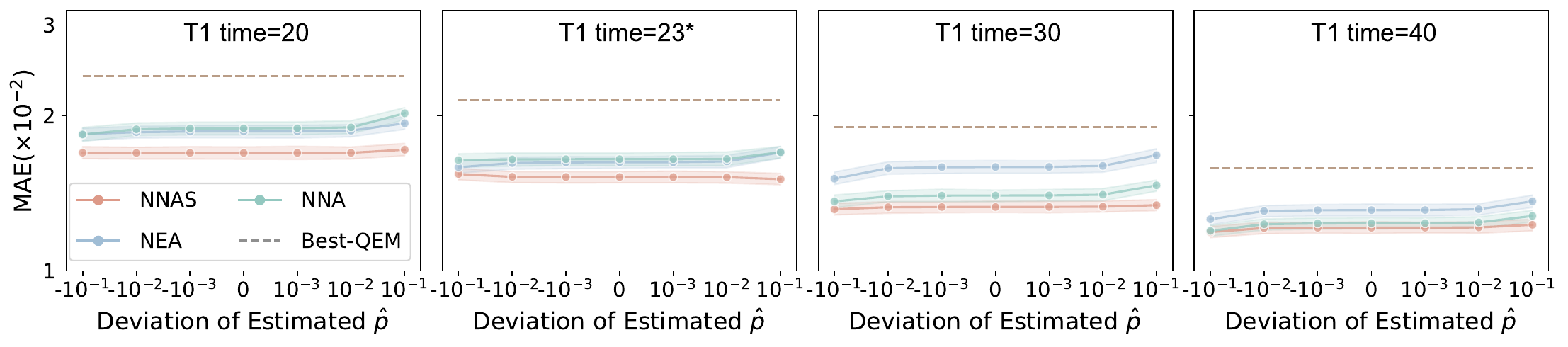}} 
    \caption{{\textbf{The performance of our model and the ablated models using estiamted value $\hat{p}$ with varying degrees of deviation from the exact value of $p$, across varying noise levels.} Longer T1 time denotes lower noise level. We focus on T1 = 23.235 $\mu$s which represents the state-of-the-art quantum coherence times. Results indicates that the impact of the deviation pf $\hat{p}$ from the exact value on the precision of our model's performance is negligible, indicating a robustness of the estimation errors. All the data in the figure presents average results with 95\% confidence interval error bars.}} 
    \label{figsm-prior}
\end{figure}

In the main text, we compute the mitigated expectation value at the $l$-th layer by utilizing the estimated effectiveness factors $\{\hat{p}_j\}_{j=1}^l$. These factors are incorporated within the maximum noise as detailed in the 'Methods' section of the main text, serving as thresholds to preserve the Completely Positive and Trace-Preserving (CPTP) characteristics of the effective noise channels ${\Lambda_l}_l$. In this section, we derive the formula for estimating these effectiveness factors.

Consider an $n$-qubit noisy layer $\tilde{u}$, which is composed of $D$ noisy gates arranged as $\tilde{u} = \tilde{G}_D \circ \ldots \circ \tilde{G}_2 \circ \tilde{G}_1$. Under the non-correlation assumption, we consider that quantum gates acting on different qubits are free from crosstalk, ensuring the independence of their respective noise channels. Each noisy gate $\tilde{G}_d$ can be decomposed into $\tilde{G}_d=\Theta_d\circ G_d$ where $\Theta_d$ is a CPTP noise channel.  
Through maximum noise decomposition, we can derive the expression of $\Theta_d$ as $\Theta_d=(1-p^d)\mathcal{I}+p^d\Lambda^d$. To calculate the effectiveness factor $p$, we shift the gate noise $\Theta_d$ backward:

\begin{equation}
    \begin{aligned}
        \tilde{G}_{d+1}\circ \Theta_d&=\Theta_{d+1}\circ G_{d+1}\circ \Theta_d \\
        &=\Theta_{d+1}\circ(G_{d+1}\Theta_d G_{d+1}^\dagger)\circ G_d\\
        &=\Theta_{d+1}\circ(G_{d+1}[(1-p^d)\mathcal{I}+p^d\Lambda^d]G_{d+1}^\dagger)\circ G_d\\
        &\equiv \Theta_{d+1}\circ \Theta_d^{G_{d+1}}\circ G_d,
    \end{aligned}
\end{equation}

where the factor $p^d$ of the noise channel $\Theta_d^{G_d}$ twirled by $G_{d+1}$ remains invariant upon backward relocation. The effectiveness factor $p$ is then computed using

\begin{equation}\label{eq-p_compute}
    p=1-\prod_{d=1}^D(1-p^d).
\end{equation}

Since the layer can be decomposed into single or two-qubit gates through circuit compilation, the estimation of $p^d$ for each gate within the layer is feasible via quantum process tomography~\cite{emerson2007symmetrized} techniques. Specifically, for the Pauli gate noise model, $p^d$ can be efficiently estimated through randomized benchmarking~\cite{magesan2012efficient}, as we utilize in our simulations involving the Pauli noise model. Experimental results in Fig.~\ref{figsm-prior} for Quantum Approximate Optimization Algorithm (QAOA)-type circuits under stochastic Pauli noise model indicate that our model shows resilience in estimating prior value $p$, displaying a moderate level of tolerance.

\section{Quantum error mitigation methods in baselines}
In this section, we will provide the details for typical quantum error mitigation (QEM) strategies that serve as baselines in the 'Results' section of main text, along with the precise parameter settings employed in their implementation. 

\subsection{Standard quantum error mitigation methods}
The standard QEM methods we employ include the zero-noise extrapolation (ZNE), probabilistic error cancellation (PEC) and Clifford Data Regression (CDR). And for each standard QEM method, we will evaluate the samping overhead which we quantify by the total execution time for a mitigation process. This metric is given by the formula $T=N_c(1000+N_s)$~\cite{Alireza2022Shadow}, where $N_c$ is the total number of circuit instances required for mitigation, and $N_s$ is the measurement shots number per circuit. The comparative demonstration of the sampling overhead for mitigation for baselines is given in the 'Results' section of main text.

\subsubsection{Zero-Noise Extrapolation}
ZNE is a canonical QEM technique that leverages the principle of extrapolating noisy expectation values to estimate the noiseless value. This method is particularly adept at handling errors that arise from the quantum hardware itself. ZNE operates by scaling the noise levels in a controlled manner and then extrapolating the noisy results to the zero-noise limit. We hypothesize the ability to scale the noise level to integer multiples of the base noise level $\epsilon$, with the multiples $1=x_0<x_1<...<x_r$. We execute the circuits under these amplified noise levels $\{x_i\epsilon\}_{i=0}^r$ and collect the corresponding noisy expectation values $\{\langle \tilde{O}\rangle_i\}_i$. Next, we model the expectation values as a function of $\epsilon$, and then extrapolate this function to estimate the noise-free limit.
The mitigated expectation value is calculated using Richardson extrapolation, which employs a linear equation to extrapolate the noise-free limit. The calculation is
\begin{equation}
    \langle O\rangle{_\text{ZNE}}=\sum_{i=0}^r\gamma_i\langle \tilde{O}\rangle_i,
\end{equation}
where the fitting coefficients $\{\gamma_i\}_i$ meet $\sum_{i=0}^r\gamma_i=1$ and $\sum_{i=0}^r\gamma_ix_i^j=0$ for $j=1,...,r$ and the solutions gives 
\begin{equation}
    \gamma_i=\prod_{i\neq j}\frac{x_j}{x_j-x_i}.
\end{equation}
We implement two-point noise amplification and apply Gate Unfolding~\cite{Endo2018Practical} as our method for noise scaling. Specifically, within the noise-amplified setup, each multi-qubit gate in the circuit is applied $2m+1$ times, with $2m+1$ representing the amplification factor. We have chosen $m=1$, corresponding to a threefold amplification of the noise level. In our setting, the implementation of ZNE involves two circuit instances: one for the circuit under the original noise level and another for the circuit with noise amplified threefold. Each instance is run with $N_s=8192$ measurement shots. Consequently, we require a total of $8192 \times 2 = 16384$ shots to achieve a comprehensive mitigation process, resulting in the sampling overhead for mitigation $T=2\times(1000+8192)=18384$.

\subsubsection{Probabilistic Error Cancellation}
PEC is one of the earliest QEM algorithms proposed. Its fundamental concept involves calibrating the noisy process channel expression to implement the inverse of the noise channel, which is then appended to the noisy circuit. We utilize the PEC method with sparse Pauli-Lindblad models (SPL)~\cite{vandenBerg2023Probabilistic,Jaloveckas2023EfficientLS}. Given that single-qubit Pauli gates, which are intended for noise cancellation, are themselves susceptible to noise contamination, their effectiveness in mitigating noise for single-qubit gates is limited. In contrast, the noise impact of two-qubit gates is significantly greater than that of single-qubit gates. Consequently, the additional noise introduced by single-qubit gates is negligible compared to the noise reduction achieved by error mitigation for two-qubit gates. Based on this analysis, our focus is primarily on mitigating the impact of noisy two-qubit gates. For a two-qubit noisy gate, the SPL model is expressed as:
\begin{equation}
    \mathcal{E}(\cdot) = \prod_k \left( \omega_k \mathcal{I} + (1 - \omega_k) \mathcal{P}_k \right),
\end{equation}
where $\omega_k = \frac{1 + e^{-2\lambda_k}}{2}$ and $\{\lambda_k\}_k$ are the noise coefficients of the SPL model, $k$ ranges from 0 to $4^2-1$. Then we need to derive the inverse expression of $\mathcal{E}$ to implement PEC. This derivation is notably efficient, as it simply involves negating the noise coefficients $\{\lambda_k\}_k$. This straightforward approach offers a significant advantage in that it circumvents the need for complex inversion of the full noise channel $\mathcal{E}$, which often leads to unphysical channels. These noise coefficients are related to the diagonal entries of the Pauli Transfer Matrix (PTM) of the SPL noise channel $\mathcal{E}$, which can be linked to the channel's impact on the Pauli operator $P_a$ with fidelity $f_a$:

\begin{equation}
    f_a = \frac{1}{2^n} \text{Tr}[P_a^\dagger \mathcal{E}(P_a)] = \frac{1}{2^n} \text{Tr}[P_a^\dagger P_a \prod_{\{a,k\}=0} (2\omega_k - 1)] = e^{-2 \sum_{\{a,k\}=0} \lambda_k},
\end{equation}

where $\{a,k\}=0$ denotes that the Pauli operators $P_a$ and $P_k$ anti-commute. We calculate the noise coefficients $\{\lambda_k\}_k$ via

\begin{equation}
    -\log(f)/2=M\lambda,
\end{equation}

where $M$ is a matrix characterizing the commutativity for Pauli operators, with the entry $M_{a,b}=0$ representing that Paulis $P_a$ and $P_b$ commute and $M_{a,b}=1$ otherwise. 
The Pauli fidelity, denoted as $f_a$, can be determined by fitting the decay function $a_0 \hat{f}_a^m$. This approach effectively mitigates the impact of state preparation and measurement (SPAM) errors. Here, $m$ denotes the number of repetitions of the noisy multi-qubit gates. In our experiment setting, the multi-qubit gate consists solely of the CNOT gate, which is self-conjugate. We select an even set of $m$ values, specifically $m = \{2, 4, 6, 8, 10\}$, to fit the decay function.

After determining the noise coefficients $\{\lambda_k\}_k $, we calculate the set $\{\omega_k\}_k$ and derive the full inverse channel using:

\begin{equation}
    \mathcal{E}^{-1}(\cdot) = \gamma \prod_k (\omega_k \mathcal{I} - (1 - \omega_k) \mathcal{P}_k),
\end{equation}

where $\gamma$ is the sampling overhead:

\begin{equation}\label{eq-gamma}
    \gamma = \prod_k (2\omega_k - 1) = e^{\left(\sum_k 2\lambda_k\right)}.
\end{equation}

We then append the inverse channel $ \mathcal{E}^{-1}$ to the noisy gate, employing the quasi-probability decomposition~\cite{vandenBerg2023Probabilistic}. For each $k$, we sample the identity $\mathcal{I}$  with probability $\omega_k$ and the Pauli channel $P_k$ with probability $1 - \omega_k$, and subsequently append it to the noisy gate. We keep track of the sign $(-1)$ each time a Pauli channel is sampled. This procedure is repeated for each two-qubit gate in the circuit, and all the signs accumulate. Finally, we adjust the outcomes by multiplying $(-1)^m\gamma = (-1)^m\prod_{j=1}^G \gamma_i$ across $G$ two-qubit gates, where $m$ is the number of chosen Pauli channels and $\gamma_i$ is the sampling overhead for each gate, calculated using Eq.~\ref{eq-gamma}. In our setting, we sample 100 random circuit instances for a complete mitigation process, with a total measurement shots number $N_u\times N_s=2\times8192=16384$. Then the sampling overhead for mitigation for PEC with SPL model is $T=N_u(1000+N_s)=1000N_u+N_u\times N_s=100\times 1000+16384=1.16384\times 10^5$.

\subsubsection{Clifford Data Regression}
CDR is a typical learning-based QEM method, which can be regarded as the simplest form of linear model-based machine learning quantum error mitigation (ML-QEM) method. CDR differs from methods like PEC, which depend on exact knowledge of the noise channel, and ZNE, which requires precise control over the noise level. Instead, CDR estimates the expectation values of target circuits by establishing a linear relationship between noisy and noiseless expectation values. This relationship is learned using training datasets derived from near-Clifford circuits that are relevant to the target circuit. These near-Clifford circuits, which are classically simulatable, are constructed by substituting some non-Clifford gates in the target circuit with Clifford gates. 

Our training dataset comprises near-Clifford circuits with a replacement rate of $r_p = 0.5$. The non-Clifford gates we utilized include the X-rotation gate $R_x(\theta)$, the Y-rotation gate $R_y$, and the Z-rotation gate $R_z$. We replace a proportion of $r_p$ of these gates with single-qubit Clifford gates according to the following replacement rule,

\begin{equation}
    \begin{aligned}
        &R_x(\theta)\Longrightarrow R_x(\frac{k\pi}{2}), \\
        &R_y(\theta)\Longrightarrow R_y(\frac{k\pi}{2}), \\
        &R_z(\theta)\Longrightarrow R_z(\frac{k\pi}{2}),
    \end{aligned}
\end{equation}

where $k$ is randomly sampled with the weight computed by

\begin{equation}
    w(k)=e^{-4d^2}, d=\|R_{x,y,z}(\theta)-R_{x,y,z}(\frac{k\pi}{2})\|.
\end{equation}

For each target circuit, we generate and measure 10 corresponding near-Clifford circuits for training, each containing a maximum of 20 non-Clifford gates. These measurements are conducted using the identical observable as that of the target circuit, with each circuit instance receiving 8192 measurement shots. Consequently, the total number of measurement shots required for the complete mitigation process using CDR is calculated as $8192 \times 11 = 90112$. Then the sampling overhead for mitigation of CDR is $T=11*(1000+8192)=1.01112\times 10^5$.

\subsection{Data-driven machine learning quantum error mitigation methods}
We choose the typical data-driven machine learning model random forest to mitigate noisy circuits, and we refer to this approach as RF. RF is an ensemble machine learning method that integrates the prediction of multiple decision trees. Comparing with a single decision tree, the RF method exhibits a preferable capacity against over-fitting, but it also entails greater consumption of computing resources. We utilize the random forest regression algorithm following~\cite{ML4QEM} for QEM, with the RandomForestRegressor as a RF implement and the GridSearchCV as an exhaustive parameter search. The setting of parameter searching grid are as table \ref{grid-search-setting}. 

\begin{table}
\caption{The setting of grid search.}\label{grid-search-setting}%
\begin{tabular*}{\columnwidth}{@{\extracolsep{\fill}}cc|ccc|cc|cc|cc@{}}
\toprule
\multicolumn{2}{c|}{number of estimators} & \multicolumn{3}{c|}{maximum depth} & \multicolumn{2}{c|}{minimum samples split} & \multicolumn{2}{c|}{minimum samples leaf} & \multicolumn{2}{c}{maximum features} \\
\midrule
min & max & option & min & max & min & max & min & max & number & option\\
\midrule
 100 & 200 &  None & 10 & 20   & 2 & 5  & 1 & 2 & 1 & sqrt \\
\bottomrule
\end{tabular*}
\end{table}

\subsection{Our models including ablation models}
In the main text, we use two ablation models: the neural extractor ablated (NEA) and the quantum-inspired
neural network ablated (NNA). The two models, NEA and NNA, are both founded on the principle of isolating the noise envelope $R_l$, aiming to the estimated noise impact factor $\hat{r}_l$. For all ML model-based baselines, including RF, our model, NEA, and NNA, the samping overhead for mitigation matches that of the noisy case (Noisy), totaling 9192. This is because each circuit instance is measured with 8192 shots, leading to a calculation of $1000 + 8192 = 9192$.

\subsubsection{Attention-core ablated model}
The architecture of the NEA model is rooted in the neural noise accumulator, which is realized by a Recurrent Neural Network (RNN).
In our model, we integrate circuit features and concatenate them to form the input data matrix $X \in \mathbb{R}^{L \times M}$, where $L$ denotes the sequence length, and $M$ represents the number of variables. The RNN within our model consists of a hidden state $H$ that captures information from prior inputs, and an output $Y$, which is a sequence of the predicted results. At each time step $l$, the hidden state $H_l$ updates recursively according to the following equation:
\begin{equation}
    H_l = f(H_{l-1}, X_l) = \tanh(W_1 H_{l-1} + b_1 + W_2 X_l + b_2),
\end{equation}
where $W_1, b_1, W_2, b_2$ are learnable parameters. Subsequently, a readout layer, consisting of a linear transformation, is employed to extract the noise component $Y_l$:
\begin{equation}
    Y_l = g(H_l) = W_3 H_l + b_3,
\end{equation}
with $W_3, b_3$ being additional learnable parameters. Ultimately, the desired output $Y_l^*$ is computed as:
\begin{equation}
    Y_l^* = \frac{\langle \tilde{O} \rangle_l}{\prod_{j=1}^l (1 - \hat{p}_j) + Y_l}.
\end{equation}
Here, $Y_l^*$ represents the predicted results.

\subsubsection{Quantum-inspired
neural network ablated}
We then explore the ablation of the entire physics-inspired neural architecture by employing a standard artificial intelligence approach: the Multilayer Perceptron (MLP). The MLP utilized in the NNA consists of multiple hidden layers, each defined by the following linear transformation,
\begin{equation}
\label{linear-equation}
    X^{(i)} = W^{(i)} X^{(i-1)} + b^{(i)},
\end{equation}
where $X^{(i)} \in \mathbb{R}^{L \times M}$ represents the output of the $i$-th layer, $W^{(i)}$ is the learnable weight matrix for the $i$-th layer, and $i$ ranges from 1 to 5. In our experiments, the MLP comprises 5 layers, with each hidden layer having a size of 32 units. The input to the first layer, $X^0$, consists of embedded features $X \in \mathbb{R}^{L \times D}$ derived from the original input data, which includes device and circuit specifications along with noisy results. The final output of the network, $Y \in \mathbb{R}^{L \times 1}$, is produced by the last layer. Here, $L$ denotes the length of the sequence, and $D$ represents the dimensionality of the input features.

\section{Dataset generation}
The dataset utilized in the main text originates from noisy simulations conducted using Qiskit~\cite{Qiskit2024qiskit}. We will provide a detailed exposition of the dataset's generation process and its key characteristics.

\subsection{Gate noise model used for simulation}
In our numerical simulations to generate datasets, we adopt a noise model derived from the principles of decoherence in superconducting systems, a prevalent architecture in quantum computing. Specifically, we identify amplitude damping (AD) and phase damping (PD) as the primary sources of decoherence, which are commonly characterized by the system's relaxation time $T_1$ and dephasing time $T_2$~\cite{Ghosh2012Surface,Tomita2014Low}. For a single qubit, the AD channel has the form
\begin{equation}
    \mathcal{E}_\text{AD}(\rho)=E_1^\text{AD}\rho E_1^{\text{AD}^\dagger}+E_2^\text{AD}\rho E_2^{\text{AD}^\dagger},
\end{equation}
where $E_1^\text{AD}$ and $E_2^\text{AD}$ are defined as
\begin{equation}
    E_1^\text{AD}=\begin{pmatrix}
1 & 0 \\ 0 & 
\sqrt{1-p_\text{AD}}\end{pmatrix} \ \text{and} \ E_2^\text{AD}=\begin{pmatrix}
0 & \sqrt{p_\text{AD}} \\ 0 & 
0\end{pmatrix}
\end{equation}
which are the Kraus operators of the AD noise channel $\mathcal{E}_\text{AD}$ with error rate $p_\text{AD}$. As for PD noise channel, the Kraus operators $E_1^\text{PD}$ and $E_2^\text{PD}$ are
\begin{equation}
    E_1^\text{PD}=\begin{pmatrix}
1 & 0 \\ 0 & 
\sqrt{1-p_\text{PD}}\end{pmatrix} \ \text{and} \ E_2^\text{PD}=\begin{pmatrix}
0 & \sqrt{p_\text{PD}} \\ 0 & 
0\end{pmatrix}
\end{equation}

To effectively simulate these sources of noise, one can employ the randomized compiling (RC)~\cite{Wallman2016Noisetailoring,Erhard2019Characterizing,Hashim2021Randomizedcompiling} technique. The RC technique is utilized to map the complex effects of AD and RD directly onto a more computationally tractable Pauli noise channel. 
In our case, we adopt the Pauli noise model as a direct outcome of applying the RC technique to the AD and PD noise channels, which can be 
\begin{equation*}
    \begin{aligned}
        \mathcal{E}^\text{P}(\rho)=(1-p_X-p_Y-p_Z)\rho 
        +p_XX\rho X+p_YY\rho Y+p_ZZ\rho Z,
    \end{aligned}
\end{equation*}
where the parameters of the Pauli noise model are determined by the decoherence rates $T_1$ and $T_2$:
\begin{equation*}
    \begin{aligned}
        p_X=p_Y=\frac{1-e^{-t/T_1}}{4} \ \text{and} \ 
        p_Z=\frac{1-e^{-t/T_2}}{2}-\frac{1-e^{-t/T_1}}{4}.
    \end{aligned}    
\end{equation*}
We employ the average $T_1=23.2357\mu$s and $T_2=15.6\mu$s from Ref~\cite{Morvan2024Phase} as a baseline for scaling the noise proportionally, while maintaining the ratio between them. 
Under the assumption of uncorrelated errors, the Pauli error probabilities for two-qubit Pauli noise channel can be quantified as follows~\cite{Ghosh2012Surface}:
\begin{equation*}
    \begin{aligned}
        &p_{IX}=p_{XI}=p_{IY}=p_{YI}=p_X(1-p_X-p_Y-p_Z), \\
        &p_{XX}=p_{XY}=p_{YY}=p_{YX}=p_Xp_Y, \\
        &p_{XZ}=p_{ZX}=p_{YZ}=p_{ZY}=p_Xp_Z, \\
        &p_{IZ}=p_{ZI}=p_Z(1-p_X-p_Y-p_Z), \\
        &p_{ZZ}=p_Zp_Z.
    \end{aligned}
\end{equation*}
For the gate execution time $t$, we set the single-qubit gate time to 18 ns and the two-qubit gate time to 48 ns, corresponding to their respective Pauli error rates. Moreover, we have incorporated a fluctuation of $2\times 10^{-5}$ to emulate the intrinsic numerical variations observed in actual quantum computing environments. The Pauli error rate, which is the aggregate of all individual Pauli error probabilities, across varying noise levels is detailed in Table \ref{tabsm-Perror}.

\begin{table}[h]\label{tabsm-Perror}
\centering
\caption{Pauli error rates across different noise levels.}
\begin{longtable}{>{\centering\hspace{0pt}}m{0.088\linewidth}>{\centering\hspace{0pt}}m{0.088\linewidth}>{\centering\hspace{0pt}}m{0.396\linewidth}>{\centering\arraybackslash\hspace{0pt}}m{0.365\linewidth}} 
\hline
\multicolumn{2}{>{\centering\hspace{0pt}}m{0.176\linewidth}}{noise level} & \multirow{2}{*}{\Centering{}single-qubit Pauli error rate\footnotemark[1]} & \multirow{2}{*}{\Centering{}two-qubit Pauli error rate\footnotemark[2]} \\* 
\cline{1-2}
T1 time & T2 time &  &  \endfirsthead 
\hline
20.00 & 13.43 & 0.2384\% & 0.4761\% \\
23.24 & 15.60 & 0.2052\% & 0.4100\% \\
30.00 & 20.14 & 0.1590\% & 0.3177\% \\
40.00 & 26.86 & 0.1193\% & 0.2384\% \\
\hline
\end{longtable}
\footnotetext[1]{is calculated by 1-$p_X-p_Y-p_Z$.}
\footnotetext[2]{is calculated by 1-$(1-p_X-p_Y-p_Z)^2$ under the uncorrelated noise assumption.}
\end{table}

\subsection{Metric for dataset acquisition difficulty}\label{sec-pr}
In the main text, we denote the \textit{hard regime} for training dataset which represents the data points obtained from deeper circuits. In the quantum sequential task, the \textit{hard regime} can be characterized by the maximum sequence length $L$ for each circuit parameter. The maximum length $L$ is determined by the selection rates listed in Table~\ref{tabsm-task1L} for QAOA-like circuit tasks and Table~\ref{tabsm-task2L} for quantum metrology, which are influenced by the partial training rate $p_r$. Consequently, the number of data points incorporated into the training set can be calculated with $L$ and the training data size $M$ which is the number of chosen set of circuit parameters. For instance, the number of data points for QAOA-type circuit tasks within a 100 training size training set and $p_r=0.25$ is $100\cdot10\cdot0.75+100\cdot13\cdot0.25\cdot0.5+100\cdot17\cdot0.25\cdot0.3+100\cdot20\cdot0.25\cdot0.2=1140$. 

\begin{table}
\caption{The selection rate for maximum Trotter step $L$.}\label{tabsm-task1L}%
\begin{tabular*}{\columnwidth}{@{\extracolsep{\fill}}llll@{}}
\toprule
10 & 11-13  & 14-17 & 18-20\\
\midrule
1-$p_r$\footnotemark[1] & $p_r\cdot 0.5$   & $p_r\cdot 0.3$  & $p_r\cdot 0.2$\\
\bottomrule
\end{tabular*}
\footnotetext[1]{denotes the partial training rate.}
\end{table}

\begin{table}
\caption{The selection rate for maximum number of qubits $n_{\text{max}}$.}\label{tabsm-task2L}%
\begin{tabular*}{\columnwidth}{@{\extracolsep{\fill}}lll@{}}
\toprule
5 & 6-8  & 9-10 \\
\midrule
1-$p_r$ & $p_r\cdot 0.7$   & $p_r\cdot 0.3$  \\
\bottomrule
\end{tabular*}
\end{table}

\subsection{Experimental settings for quantum algorithms based on the Quantum Approximate Optimization Algorithm-type circuits}

\begin{figure}[t]
    \centering
    {\includegraphics[width=0.6\textwidth]{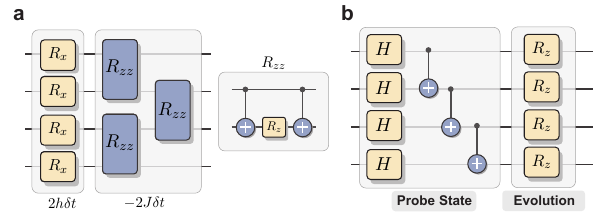}}    
    \caption{{\textbf{The quantum circuit for structured circuits used in numerical simulations.} \textbf{a} The Trotterized circuit layer with parameters $(J,h)$, which comprises a column of X-rotation gates at $2h\delta t$ and a chain of $R_{ZZ}$ gates. The right subfigure shows the $R_{ZZ}$ gate's decomposition into two CNOT gates and a Z-rotation gate at $-2J\delta t$. \textbf{b} The circuit used for quantum metrology, which includes the GHZ probe state preparation circuit and the evolution circuit. The evolution circuit $U(\theta)$ parameterized by $\theta$ consists of a column of Z-rotation gates at $\theta$. }} 
    \label{figsm-circ}
\end{figure}

The QAOA-type circuits are commonly utilized in near-term quantum algorithms such as quantum Trotterized simulation and variational quantum algorithms (VQAs).  
In the main text, we employ the Hamiltonian which determines the temporal evolution of the 1D transverse-field Ising spin chain. The Hamiltonian is,
\begin{equation}\label{eq-hising}
        H_\text{Ising}=-J\sum_{j=0}^{n-2}Z_jZ_{j+1}+h\sum_{j=0}^{n-1}X_j\equiv H^Z_\text{Ising}+H^X_\text{Ising},
\end{equation}
here we define $H^Z_\text{Ising}=-J\sum_jZ_jZ_{j+1}$ and $H^X_\text{Ising}=\sum_jX_j$. $J$ and $h$ are the parameters which denote the coupling strength between neighboring spins and the transverse magnetic field, respectively. For a certain pair of circuit parameters ($J,h$), we measure the Trotter circuit $U^{J,h}_l$ at each step $l$ ($l=1,2,...,L$) with single Z observable $Z_n$ on the last $n$-th qubit. To implement the Trotter circuit $U^{J,h}_l$, we employ the first-order Trotterized decomposition to approximate the time-evolution operator,
\begin{equation}
        e^{-iH_\text{Ising}t}=e^{-i(H^Z_\text{Ising}+H^X_\text{Ising})t}\approx e^{-iH^Z_\text{Ising}t}e^{-iH^X_\text{Ising}t}.
\end{equation}
For one time step $\delta t$, we have 
\begin{equation}
    \begin{aligned}
        &e^{-iH^Z_\text{Ising}\delta t}=\prod_je^{-i(-J\delta t)\sigma_z^j\sigma_z^{j+1}}\\
        &e^{-iH^X_\text{Ising}\delta t}=\prod_ie^{-i(h\delta t)\sigma_x^j},
    \end{aligned}
\end{equation}
where the circuit structure of a single layer is shown in Fig.~\ref{figsm-circ} \textbf{a}. 
The circuits span Trotter steps from 1 to 20. We maintain a fixed ratio of $J$ to $h$ at 0.6 and sample $h\delta t$ across the interval $[0.5, 2]$.

\begin{table}
\caption{Dataset characteristics}\label{setting-dataset}
\begin{tabular*}{\columnwidth}{@{\extracolsep{\fill}}cccccccc@{}}
\toprule 
\multirow{2}{*}{Task} & \multicolumn{2}{c}{Training Set} & \multicolumn{2}{c}{Test Set} & \multirow{2}{*}{Quantum System Size}\\
\cline{2-3}
\cline{4-5}
& sequences & hard regime & sequences & steps \\
\midrule
QAOA-type circuits & 50-500 & 11-20 & 200 & 1-20 & 6 \\
Qubit-wise circuits for quantum metrology & 100-500 & 5-10 & 500 & 1-10 & 1-10 \\
 
\bottomrule
\end{tabular*}
\end{table}

\begin{figure}[!h]
    \centering
    {\includegraphics[width=0.8\textwidth]{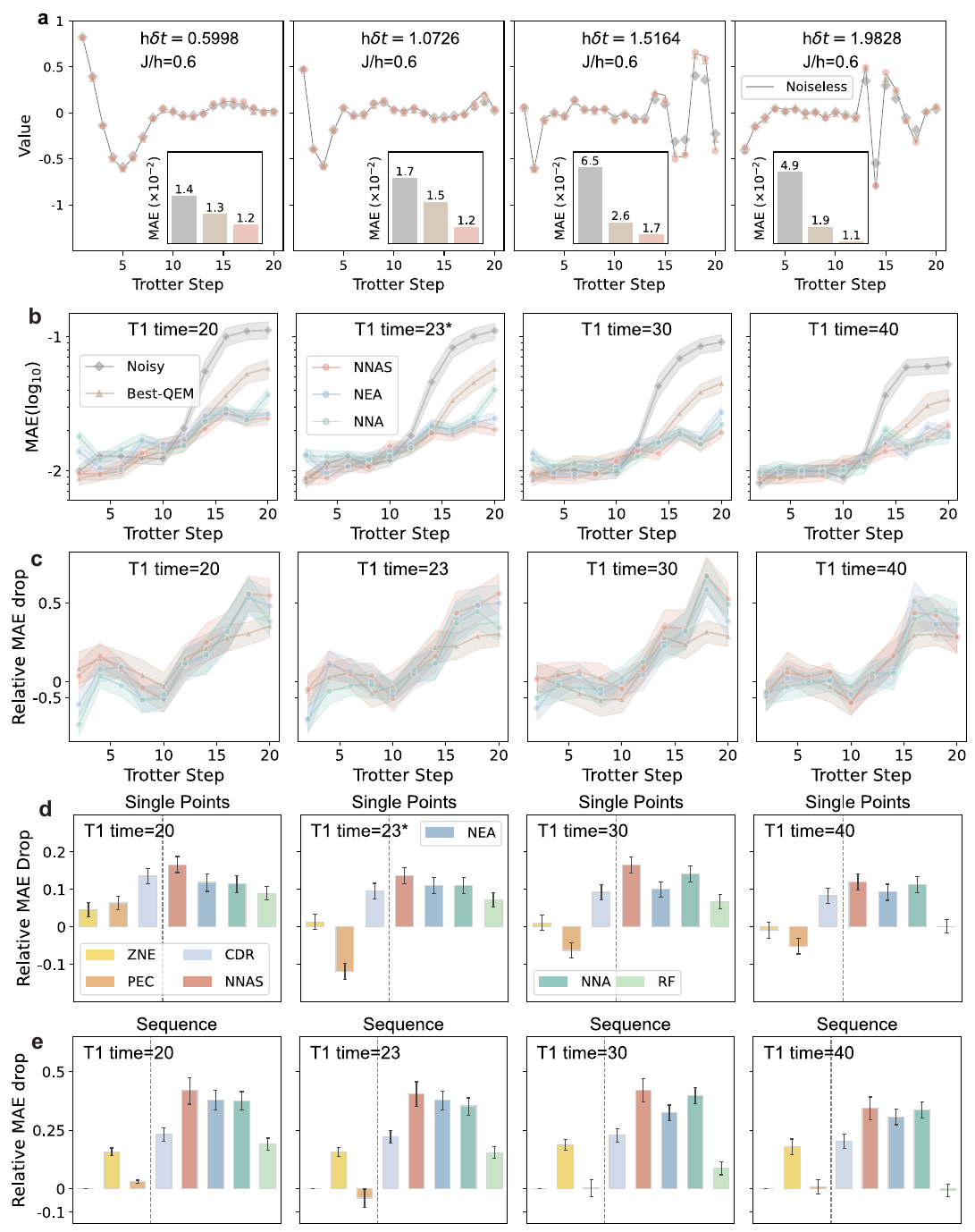}}    
    \caption{{\textbf{Extended data on accuracy analysis for Quantum Approximate Optimization Algorithm-like circuits.} \textbf{a} The Trotter evolution for typical values of circuit parameter $h\delta t$ within $[0.5,2]$. The inset figure shows the total MAE comparison for Trotter steps 1-20 among Noisy, Best-QEM (the best-performing QEM method among standard QEM methods and RF), and our model. \textbf{b}-\textbf{c} The comparative results of MAE and the relative MAE drop (RD) for baselines and our model with varying Trotter steps, where MAE and RD are both calculated using single points. These results are derived under diverse noise levels, characterized by the T1 relaxation time. A higher T1 time signifies a lower noise level. \textbf{d}-\textbf{e} The comparative analysis of RD for baselines calculated using single points and sequence norms. 
    All the data in the figure presents average results with 95\% confidence interval error bars.}} 
    \label{figsm-qs_benchmark}
\end{figure}

\begin{figure}[!h]
    \centering
    {\includegraphics[width=0.75\textwidth]{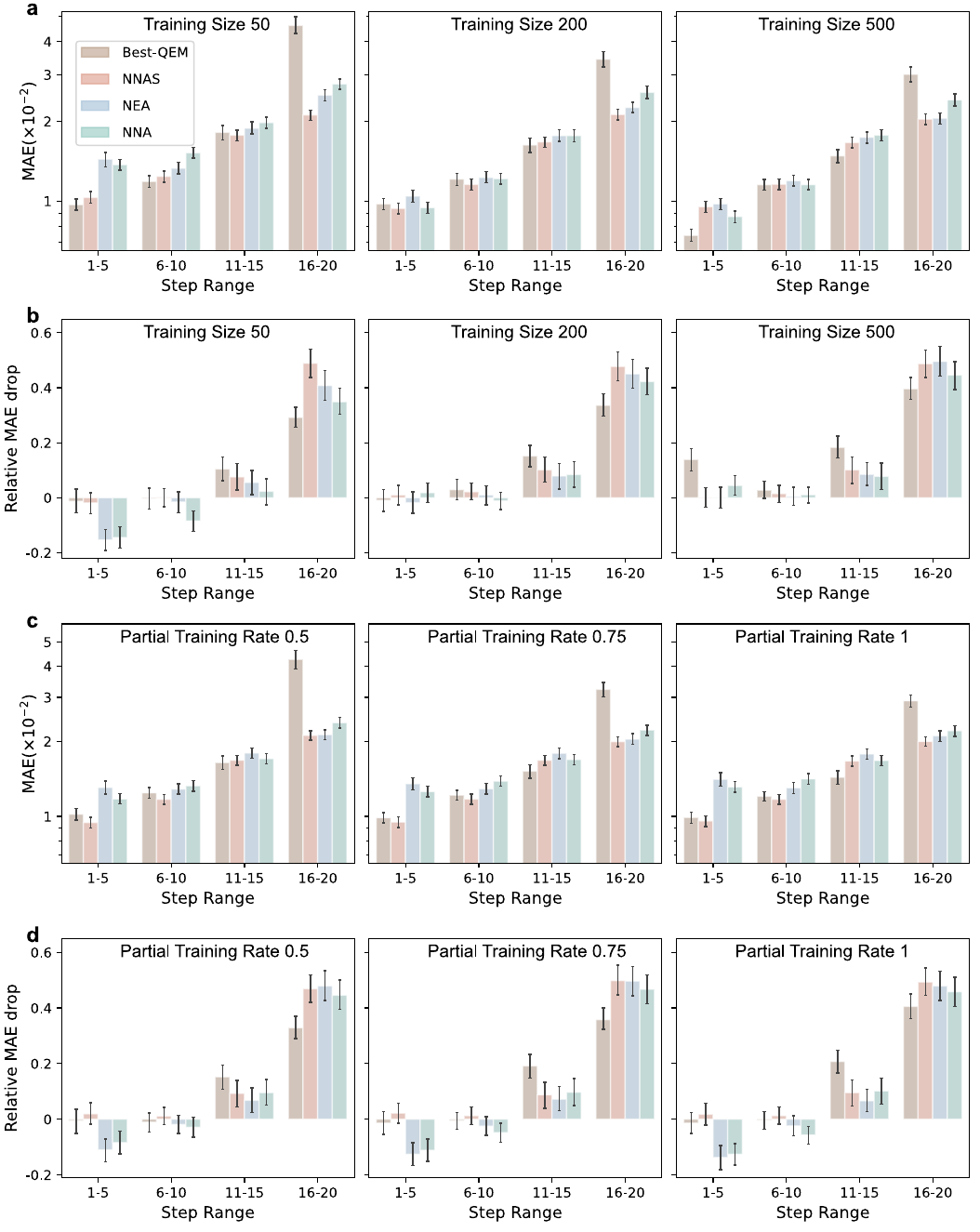}}    
    \caption{{\textbf{Extended data on efficiency analysis of ML-QEM models for Quantum Approximate Optimization Algorithm-like circuits.} The average performance for MAE and RD is assessed with varying training set size and partial training rate across specific Trotter step ranges: 1-5, 6-10, 11-15, and 16-20 (The model's performance with a training set size 100 and a partial training rate 0.25 is discussed in the main text.) All the data in the figure presents average results with 95\% confidence interval error bars.}} 
    \label{figsm-qs_efficiency}
\end{figure}

\begin{figure}[!h]
    \centering
    {\includegraphics[width=0.8\textwidth]{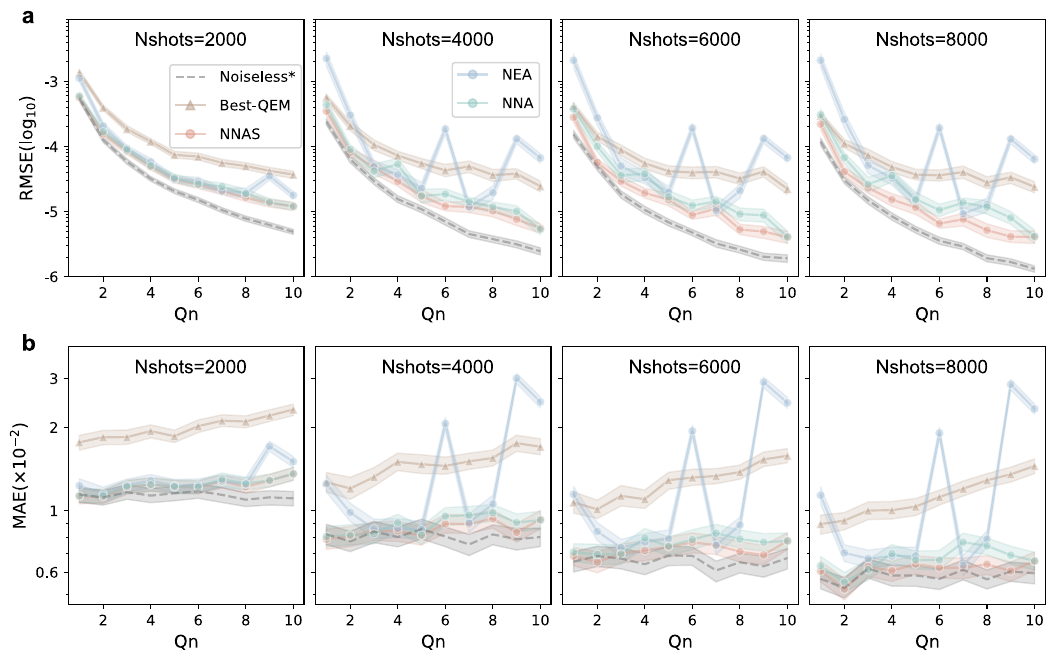}}
    \caption{{\textbf{Extended data on accuracy analysis for quantum metrology based on GHZ state.} \textbf{a}-\textbf{b} The average performance of the baselines is assessed using two metrics including RMSE for estimated angles and MAE for mitigated expectation values, at each qubit number across an increasing range of measurement shots. Here, the noiseless case refers to the condition that the results only suffer measurement statistical error due to finite measurement shots number. All the data in the figure presents average results with 95\% confidence interval error bars.}}
    \label{figsm-qm_benchmark}
\end{figure}

\begin{figure}[!h]
    \centering
    {\includegraphics[width=0.8\textwidth]{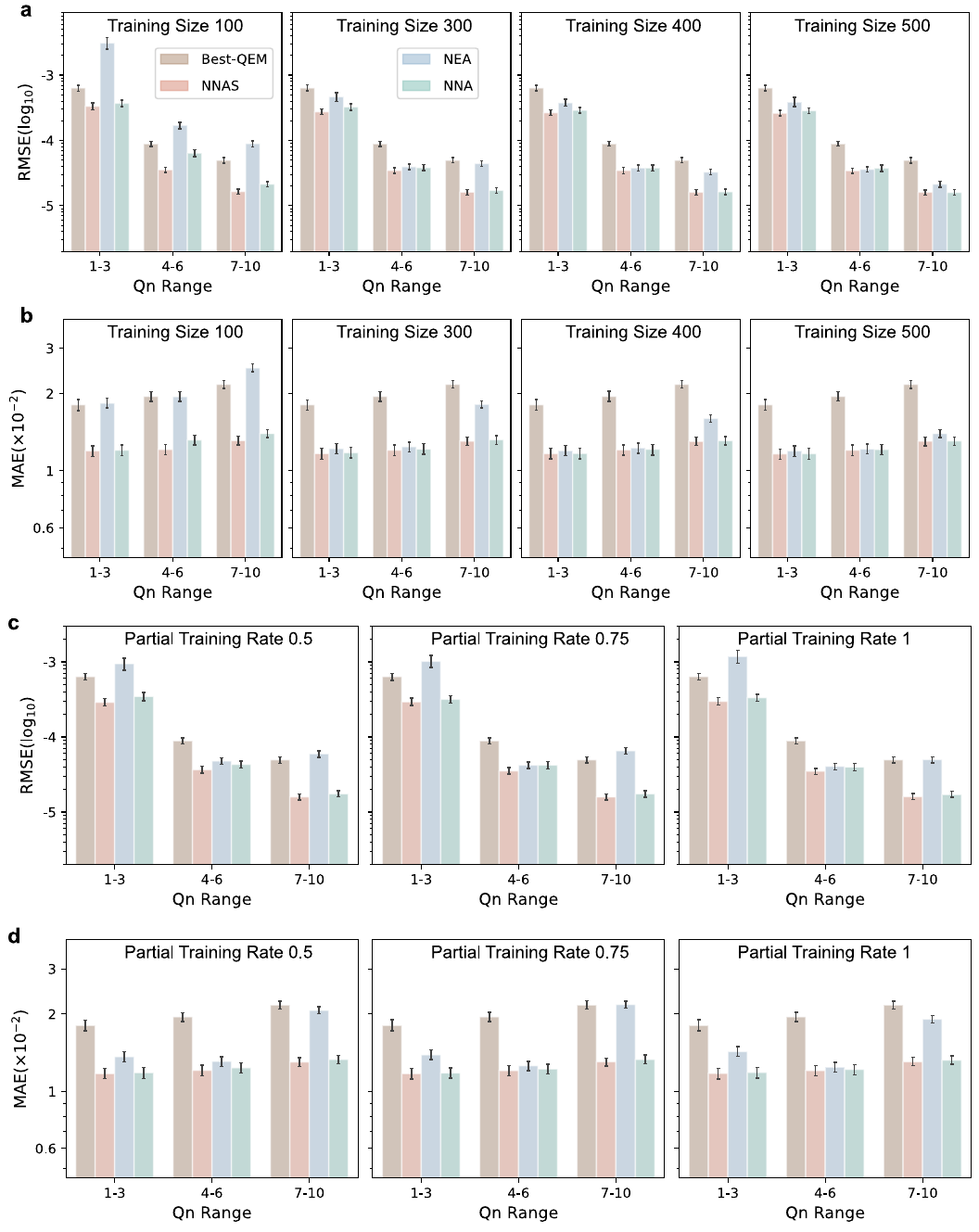}}    
    \caption{{\textbf{Extended data on efficiency analysis for quantum metrology based on GHZ state.} The average performance of RMSE for estimated angles and MAE for mitigated expectation values is assessed with varying training set size and partial training rate across specific qubit number ranges: 1-3, 4-6, and 7-10 (The model's performance with a training set size 200 and a partial training rate 0.25 is discussed in the main text.) All the data in the figure presents average results with 95\% confidence interval error bars..}} 
    \label{figsm-qm_efficiency}
\end{figure}

\subsection{Experimental settings for quantum metrology based on GHZ probe state}
Quantum metrology (QM) is one of the most appealing applications of quantum techniques, which untilizes quantum resources to estimate an unknown parameter in higher precision than the classical approaches. We focus on the QM with GHZ probe state, which enables the Heisenberg scaling $\mathcal{O}(n^{-2})$ of precision with regard to the system size $n$, while the presicion scales as $\mathcal{O}(n^{-1})$ for seperate states which called shot-noise scaling. We expose the $n$-qubit GHZ probe state  $|\text{GHZ}_n\rangle = \frac{1}{\sqrt{2}}\left(|0\rangle^{\otimes n}+|1\rangle^{\otimes n}\right)$ to the interaction dynamics which is encoded by an unknown parameter $\theta$. We consider the interaction dynamics represented by the time evolution $U(\theta)=e^{-iH\theta}$ with the Zeeman Hamiltonian $H=\frac{1}{2}\sum_{i=1}^nZ_i$. Then given the global X observable $O_n=X^{\otimes n}$ and the measurement shot number $N$, the parameter $\theta$ can be estimated via
\begin{equation}
    \theta^\text{est}(n,N)=\frac{\arccos{\langle O_n\rangle_\theta}}{n},
\end{equation}
where $\langle O_n\rangle_\theta$ refers to the expectation value of $O_n$. The circuit we utilized includes the GHZ state preparation circuit and the free evolution circuit parameterized by $\theta$, as depicted in Fig.~\ref{figsm-circ} \textbf{b}.
We train the models with circuit parameterized by randomly sampled angles within a continuous interval from 0.1 to 5.0 with increments of 0.1, performing 10 independent experimental trials to estimate metrics for each parameter in simulations.  Evaluations are based on angles taken from 0.05 to 5.05 at 0.1 increments.
 
\subsection{Summary of Dataset characteristics}
In conclusion, we summarize the dataset utilized in our experiments for two type of circuits.
The overall characteristics of the dataset are given in Table~\ref{setting-dataset}. For the first two sequential tasks in our simulation experiments, the dataset comprises sequences with varying lengths in the training set, which are randomly truncated, and sequences with fixed lengths in the test set. 
 
 

\section{Experimental details and extended results}
In this section, we detail the implementation of our experiments and present extended results.

\subsection{Experiments for QAOA-type circuits}
For layer-wise QAOA-type circuits, we initially select representative results from the interval $[0.5,2]$ of the circuit parameter $h\delta t$ to demonstrate the noiseless expectation values of the Trotter evolution. As observed in Fig.~\ref{figsm-qs_benchmark} \textbf{a}, the absolute value of these noiseless expectation values increases with the $h\delta t$ value, particularly at higher layer depths, where our model's superiority becomes increasingly evident. 

In the main text, we assessed our model's performance using Mean Absolute Error (MAE), a key metric for quantifying the precision of QEM methods. Results as given in Fig.~\ref{figsm-qs_benchmark} \textbf{b} substantiate the superiority of our model in two pivotal aspects. The first is that, as analyzed in the main text, our model demonstrates superior performance with larger Trotter steps. The second is that our model exhibits enhanced stability relative to our ablation models, notably at lower Trotter steps where the ablation models exhibit instability. This stability underscores the robustness of our model's design, which sustains effective mitigation ability as validated by ablation studies.

We then extend our analysis with an additional metric to further investigate the model's performance. We define the relative MAE reduction (RD) as a metric to quantify the relative performance improvement of mitigation methods over the noisy baseline, assessing this for each layer individually. The RD is calculated using the logarithmic ratio of the MAE for the noisy and mitigated expectation values:
\begin{equation}
    \text{RD}=\log\left(\frac{|\tilde{y}_l-y_l|}{|y_l^\text{em}-y_l|}\right),
\end{equation}
where $\tilde{y}_l$ denotes the noisy expectation value at $l$-th layer, $y_l$ represents the noiseless expectation value, and $y_l^{\text{em}}$ is the mitigated result obtained from QEM methods. Notably, a larger RD signifies better performance, indicating a more effective mitigation of noise.
Fig.~\ref{figsm-qs_benchmark} \textbf{c} depicts the RD of baselines as a function of Trotter Step, where our model shows an overall stable advantage over the baselines.
The comparative analysis of RD for baselines under different noise levels is given in Fig.~\ref{figsm-qs_benchmark} \textbf{d} and \textbf{e}, using two calculations for single points and sequences. 

We also evaluate the efficiency of our model in comparison with baselines using ML models, including RF, and our ablated models. As depicted in Fig.~\ref{figsm-qs_efficiency}, the average performance for MAE and RD is assessed across specific Trotter step ranges: 1-5, 6-10, 11-15, and 16-20. This assessment is conducted for two different conditions: training set sizes ranging from 50 to 500 as shown in panels \textbf{a} and \textbf{b}, and partial training rates varying from 0.5 to 1 as illustrated in panels \textbf{c} and \textbf{d} (The model's performance with a training set size 100 and a partial training rate 0.25 is discussed in the main text.). Our model outperforms RF and ablation models, particularly with lower training set sizes and partial training rates, demonstrating its strength in small data scenarios.

\subsection{Discussion on Accumulation Surrogating}

We further explore how our physical-inspired network surrogates the noise accumulation process. For QAOA-type circuits with up to 12 Trotter steps, we examine a specific instance, visualizing both the quantum cumulative noise $\mathcal{N}$ and the corresponding surrogate $\hat{\mathcal{N}}$ derived from the neural network. Our visualization technique employs the Pauli Transfer Matrix (PTM) representation of $\mathcal{N}$, condensed through sum pooling, alongside the computation of the outer product $\hat{\mathcal{N}}^T\hat{\mathcal{N}}$ to represent $\hat{\mathcal{N}}$. We proceed to conduct a quantitative analysis, expanding our case study to encompass a statistical analysis that includes 200 sequences. This comprehensive approach, combining both visualization and quantification, substantiates our model's rapid detection of structural changes and its structural alignment with quantum cumulative noise. Next, we will delve into the computational details that underpin these findings.

In our experiments for six-qubit QAOA-type circuits, we confront the challenge of visualizing the immense PTM size of (4096,4006). To address this, we utilize sum pooling, which not only simplifies the data representation by reducing dimensionality but also retains critical information. We employ the sum pooling on a sub matrix of size 128, and the resulting size is (32,32). For the surrogate $\hat{\mathcal{N}}$ with an experimental size of 32, the outer product results in a square matrix of size (32,32). 

When it comes to the correlation analysis in our experiments with six-qubit QAOA-type circuits, we utilize submatrices extracted from the full PTM of $\mathcal{N}$, perform a flattening operation, and then normalize each vector before computing the Spearman correlation coefficient with the flattened and normalized vector of $\hat{\mathcal{N}}^T\hat{\mathcal{N}}$. The normalization step ensures that each feature contributes equally to the correlation analysis, mitigating the impact of scale differences. The Spearman coefficient, denoted as $c_{\text{spearman}}$, is calculated by first ranking the elements of the datasets $X$ and $Y$ after normalization, and then applying the Pearson correlation formula to these ranks. The formula for $c_{\text{spearman}}$ is given by:
\begin{equation}
    c_{\text{spearman}} = \frac{\sum_i (rg(x_i) - \bar{rg_x})(rg(y_i) - \bar{rg_y})}{\sqrt{\sum_i (rg(x_i) - \bar{rg_x})^2 \sum_i (rg(y_i) - \bar{rg_y})^2}}
\end{equation}
where $rg(x_i)$ and $rg(y_i)$ represent the rank values of $x_i$ and $y_i$ after normalization, respectively, and $\bar{rg_x}$ and $\bar{rg_y}$ are the mean rank values of $X$ and $Y$. This results in a correlation matrix with dimensions of (128,128).

Furthermore, going disordered prompts us to characterize the structural correspondence through differential entropy. We analyze the variation curves of vector $\hat{\mathcal{N}}$ and the flattened vector from the PTM of $\mathcal{N}$, with a length of $4096^2$, as the Trotter step increases. The differential entropy is calculated using the formula:
\begin{equation}
    h(X) = -\int_{-\infty}^{\infty} f(x) \log f(x) \, dx
\end{equation}
where $f(x)$ is the probability density function of the continuous random variable $X$.
For these vectors, we utilize the \texttt{scipy.stats.differential\_entropy} function with the 'auto' method, which selects an appropriate numerical approach based on the sample size. Given the lengths of our vectors, the function is likely to employ the Ebrahimi method for the shorter vector $\hat{\mathcal{N}}$ and the Vasicek method for the longer flattened PTM vector.

\subsection{Experiments for quantum metrology based on GHZ probe state}
We employ three metrics to assess the mitigation effectiveness of models for QM, which we will describe in detail.
\begin{itemize}

\item Root Mean Square Error (RMSE). The RMSE is commonly used to describe the sensitivity of $\theta^{\text{est}}(n,N)$ to the statistical error given measurement shots number $N$. 
The defination of RMSE is
\begin{equation}
  \left(\Delta \theta^{\text{RMSE}}(n,N)\right)^2=\text{Var}[\theta^{\text{est}}(n,N)]=\mathbb{E}[|\theta^{\text{est}}(n,N)-\theta|^2].
\end{equation}
In the noiseless case, $\Delta\theta^{\text{RMSE}}(n,N)$ scales at $\mathcal{O}(n^{-2})$ for all $\theta$, which reaches the Heisenberg scaling.
\item MAE of the expectation value $\langle O_n\rangle_\theta$. This metric provides a direct assessment of the efficacy of the mitigation strategies. 

\item Fitting rate $r$ of the RMSE scaling curve with respect to the qubit number $n$. We model the RMSE scaling curve using the function $b_0/n^r$, where $b_0$ is determined by fitting the RMSE scaling curve of the noiseless case with the function $b_0/n^2$. 
\end{itemize}

The RMSE and fitting rate serve as indicators for QM, with the latter being an upper-bounded metric where values closer to 2 signify better performance, as they approach the noiseless scenario. The MAE, alongside the RMSE, quantifies the error in our estimates.
Since the fitting rate $r$ is calculated by fitting the data across all the qubits, we only illustrate the detailed results for RMSE and MAE at each number of qubits under varying measurement shots numbers, as given in Fig.~\ref{figsm-qm_benchmark}. Here, the noiseless case refers to the condition that the results only suffer measurement statistical error due to finite measurement shots number. Our model demonstrates performance improvements generally in accordance with the noiseless case as measurement shots number increases. This correlation arises from the decreased measurement statistical error. In terms of efficiency analysis as depicted in Fig.~\ref{figsm-qm_efficiency}, the performance of our model aligns with the outcomes observed in prior QAOA-type circuits. This refines the conclusion of the superior effectiveness and efficiency of our model compared to standard QEM and universal ML-QEM, even for qubit-wise quantum algorithm. Notably, in this experiment, the dataset we utilize encompasses a small range of circuit parameters $\{\theta\}$ (at most 50). Relative to universal ML models that rely on directly capturing the relationship between noisy and noiseless results, learning the corresponding relationship from such a dataset is particularly challenging. This further highlights the utility of physical prior knowledge in ML-QEM under conditions of limited training set.

\bibliography{b}